\documentclass[a4paper, UKenglish, cleveref, autoref, thm-restate]{lipics-v2021}

\hideLIPIcs
\nolinenumbers

\usepackage{tikz}
\usetikzlibrary{trees,positioning,shapes,shadows,arrows}
\usepackage{hyperref}
\usepackage{array}
\usepackage{graphicx}
\usepackage{pdflscape}
\usepackage{longtable}

\newcolumntype{L}[1]{>{\raggedright\arraybackslash}p{#1}}

\title{SoK: Cryptographic Key Recovery for Cryptoasset Custody and Financial Technologies}

\titlerunning{SoK: Cryptographic Key Recovery for Financial Technologies}

\author{Francisco Javier Becerra Sanchez}
{University of Luxembourg, Luxembourg, Luxembourg}
{francisco.becerra@uni.lu}
{https://orcid.org/0009-0007-4902-6550}
{}

\author{Antonio Ken Iannillo}
{University of Luxembourg, Luxembourg, Luxembourg}
{antonioken.iannillo@uni.lu}
{https://orcid.org/0000-0001-9358-7100}
{}

\author{Radu State}
{University of Luxembourg, Luxembourg, Luxembourg}
{radu.state@uni.lu}
{https://orcid.org/0000-0002-4751-9577}
{}

\authorrunning{F.\,J. Becerra Sanchez, A.\,K. Iannillo, and R. State}

\Copyright{Francisco Javier Becerra Sanchez, Antonio Ken Iannillo, and Radu State}



\funding{This research was funded in part by the Luxembourg National
Research Fund (FNR), grant NCER22/IS/16570468/NCER-FT, and by
FinnovationHub, funded by the Ministry of Finance of the Government
of Luxembourg.}

\acknowledgements{The authors thank colleagues and reviewers for helpful feedback.}

\supplementdetails[subcategory={Reproducibility Package}]
{Collection}
{https://github.com/franciscobecerra97/Cryptographic-Key-Recovery-for-Financial-Technologies/tree/aft-2026-artifacts}

\begin{document}

\maketitle

\begin{abstract}
Cryptoasset systems often bind cryptographic key control to financial control: losing a wallet seed, custody share, hardware device, or smart-account credential can remove spend authority, while compromised recovery can enable theft. Existing work treats recovery through separate vocabularies--key backup, secret sharing, account recovery, credential re-issuance, social recovery, and asset migration--making mechanisms and tradeoffs difficult to compare.

This paper presents a Systematization of Knowledge (SoK) on cryptographic key recovery for cryptoasset custody and financial technologies. Starting from a 118-paper systematic-review discovery corpus, we derive a 77-paper synthesis corpus and code each retained system in a master matrix covering recovered objects, recovery semantics, mechanisms, enrollment and storage, authorization, trust placement, failure events, post-recovery state, validation evidence, deployment status, privacy, usability, and limitations. The matrix supports an axis-first taxonomy that separates secret-restoring, hybrid, control-restoring, forensic/extractive, and framework-oriented recovery.

Our central observation is that recovery is not a single operation: systems may reconstruct an original secret, regenerate a seed, restore a share, reissue a credential, migrate signing authority, restore account control, move assets, or extract forensic artifacts. We derive a generalized construction model, check it against production-facing designs, and identify six findings: recovery semantics are heterogeneous; recovery shifts trust; liveness improvements create abuse paths; post-recovery lifecycle management is uneven; protocol evidence outpaces user evidence; and recovery metadata remains underprotected. These gaps motivate a research agenda for recovery-aware financial technologies.
\end{abstract}

\section{Introduction}
\label{sec:introduction}

In cryptoasset systems and related financial technologies, control of a cryptographic key can become control of financial value: loss may remove access to funds or services, while compromised recovery may transfer that control to an attacker. Key recovery is therefore part of the security boundary for systems that rely on cryptographic authorization \cite{8,31,54,101}.

Recovery must balance two goals that pull against each other. It should prevent legitimate users and institutions from being permanently locked out after loss, unavailability, or operational failure. At the same time, it must not create a weaker path for theft, coercion, surveillance, insider abuse, or unauthorized transfer. This tension is the core reason recovery remains difficult: the mechanisms that improve availability also introduce new actors, records, and decisions that must be secured.

Existing work proposes many mechanisms, but uses incompatible vocabulary. Recovery may mean restoring cryptographic material, changing account control, moving value, reissuing identity credentials, repairing custody roles, or extracting forensic artifacts. Comparisons can therefore obscure what is actually restored and what risks remain after recovery.

This paper addresses that fragmentation through a Systematization of Knowledge (SoK) focused on cryptographic key recovery for cryptoasset custody and financial technologies. We center the paper on financial control, while including adjacent identity, data-protection, governance, and secure-hardware work when it contributes a recovery pattern that can transfer to financial systems.

Recovery is not one operation. A useful comparison must ask what control is restored, who authorizes it, what failure it handles, and what state remains afterward.

\subsection{Contributions} 
This Systematization of Knowledge makes the following contributions:

\begin{itemize}
    \item \textbf{Financial-technology corpus and scope.}
    We construct a 118-paper discovery corpus on key backup, recovery, and restoration, with auditable queries, screening flow, exclusion reasons, snowballing rounds, QA threshold, and manually verified LLM-assisted screening. We then derive a 77-paper synthesis corpus scoped around financial technologies while retaining transferable recovery patterns from identity, hardware, and data-protection systems.

    \item \textbf{Recovery semantics for financial control.}
    We distinguish secret-restoring recovery from control-restoring recovery and show why this distinction matters for financial systems. This clarifies when a system restores the same secret and when it instead restores financial control through a changed control relationship.

    \item \textbf{A 77-system master matrix and codebook.}
    We code every retained system using a controlled vocabulary for financial context, recovery semantics, mechanism family, enrollment and storage, failure events, authorization, trust placement, adversary model, post-recovery state, validation evidence, deployment status, privacy, usability, post-quantum relevance, and limitations. Appendix~\ref{app:master-matrix} reproduces the complete per-system matrix, while the reproducibility package provides the machine-readable matrix and its codebook.

    \item \textbf{Evaluation vocabulary, taxonomy, and construction model.}
    We organize mechanisms by whether they restore secrets, restore control, combine both, perform forensic extraction, or define requirements. We define adversary classes and recovery-specific safety, liveness, and privacy properties, then derive a seven-stage construction model.

    \item \textbf{Production practice checks and research agenda.}
    We check the taxonomy against six production or standards-facing recovery designs and tie six findings to matrix counts and examples, including sparse user-study evidence, under-specified post-recovery hygiene, metadata leakage, custody boundaries, recovery economics, inheritance, and post-quantum migration.
\end{itemize}

\subsection{SoK Framing and Novelty}
\label{subsec:sok-framing-novelty}

\paragraph*{Motivation for systematization}
The goal is not only to summarize prior mechanisms. We define comparable recovery semantics, code every retained system into the master matrix reproduced in Appendix~\ref{app:master-matrix}, and separate secret reconstruction from restored financial control. A threshold wallet that reconstructs an old seed, a smart account that rotates ownership, and a recovery service that moves assets are all recovery systems, but they leave different post-recovery states and abuse paths.

\paragraph*{Novelty relative to closest related work}
Prior work usually studies one slice of the problem: cryptocurrency protocols and DeFi, wallet protocols and attacks, threshold recovery, deterministic or hardware-wallet designs, smart-account recovery, identity recovery, or forensic extraction. This SoK contributes a cross-domain comparison centered on recovery semantics rather than mechanism names. Its novelty claims are: \emph{(i)} secret-restoring and control-restoring recovery should be evaluated separately; \emph{(ii)} post-recovery state is a first-class security property for financial systems; and \emph{(iii)} recovery evidence must connect cryptographic mechanisms with operational controls, metadata exposure, production practice, and human behavior. Table~\ref{tab:related-work-comparison} compares the closest prior-work clusters.

\begin{table}[!h]
  \centering
  \scriptsize
  \caption{Closest related-work clusters and this paper's added recovery-semantics focus.}
  \label{tab:related-work-comparison}
  \begin{tabular}{@{}p{0.15\linewidth}p{0.17\linewidth}p{0.16\linewidth}p{0.14\linewidth}p{0.16\linewidth}p{0.14\linewidth}@{}}
    \hline
    Prior work & Scope & Recovery semantics & Financial-control treatment & Corpus or evidence & This paper adds \\
    \hline
    Cryptocurrency and DeFi SoKs \cite{bonneau2015sokBitcoin,werner2022sokDeFi,zhou2023sokDefiAttacks} & Protocols, DeFi primitives, risks, and incidents & Mostly indirect: key management, incidents, rescue windows, or mitigation & Protocol and market-level financial control & SoK syntheses and incident datasets & Recovery-specific matrix for wallets, custody, smart accounts, identity, and forensics \\
    Wallet taxonomies and security surveys \cite{karantias2020taxonomyWallets,houy2023walletSlr,homoliak2024sokWallets,erinle2025sokWallets} & Wallet definitions, protocols, factors, attacks, and defenses & Recovery is a feature or fallback, not the organizing axis & Wallets as transaction interfaces and attack targets & Taxonomies, SLR evidence, and incident analysis & Recovered objects, secret/control axis, post-recovery state, and lifecycle evidence \\
    Custody and key-management work \cite{10,101} & Dynamic access structures, threshold custody, HSMs, and operations & Access update, backup restoration, share reconstruction, or signer continuity & Central: assets must stay live without weakening controls & Formal constructions, prototypes, benchmarks, and cases & Cross-family comparison with regulation, economics, inheritance, production cases, and users \\
    Web3 account recovery \cite{8,37,59,82,87,109,114} & Guardians, fallback signatures, passkeys, pre-signed transfers, and composition & Concrete control-restoring mechanisms & Direct wallet and smart-account asset control & Mechanism papers, simulations, gas costs, and composability & Shows how control restoration differs from secret restoration and exposes hygiene/metadata gaps \\
    Identity and WebAuthn recovery \cite{31,43,94,102,106,117} & SSI wallets, DID backup, backup authenticators, credential reissue, and share restoration & Credential or identity-control recovery & Transferable to regulated payments and account access & Prototypes, benchmarks, and formal analyses & Separates identity restoration from asset-control recovery and maps transfer to finance \\
    Usability and key-management studies \cite{6,15,66,108} & Human coordination, wallet mental models, memory cues, seed phrases, and factor loss & User-facing ceremonies and factor handling & Often implicit through wallet access & Three retained user-study rows plus adjacent usability evidence & Conservative user-evidence table and agenda for stress, death, phishing, and accessibility \\
    \hline
  \end{tabular}
\end{table}

\section{Background and Scope}
\label{sec:background-scope}

This paper uses the term \emph{cryptographic key recovery} broadly, but not without distinction. In the corpus, recovery may mean reconstructing an original private key, regenerating a deterministic seed, restoring a key share, reissuing a credential, transferring signing authority, recovering account control, or moving assets to a new control path. These operations differ in what they restore and in the risks they introduce.

The primary scope is financial technology: systems in which cryptographic material authorizes assets, transactions, account access, custody operations, or credentials used by financial infrastructure. This includes blockchain wallets, smart-contract accounts, custody systems, decentralized applications, decentralized identity, and secure hardware when they support financial control or transferable recovery patterns.

The paper also considers non-financial work when it contributes a mechanism relevant to financial systems. Examples include identity-management recovery, data-protection recovery, governance-related recovery, and hardware-enforced recovery. These systems are included because financial technologies often reuse the same primitives and operational patterns: secret sharing, threshold reconstruction, deterministic derivation, secure elements, trusted authorities, biometric or physical anchoring, and delegated recovery.

\section{Review Methodology and Corpus Construction}
\label{sec:methodology}
The contribution of this paper is the systematization, not the review protocol itself. We nevertheless used a systematic literature review (SLR) workflow as an auditable discovery process and then narrowed the final synthesis around recovery in cryptoasset custody and financial technologies.

\subsection{Guiding Questions}
\label{subsec:rqs}
The review was guided by four questions: \textbf{RQ1:} What object or control relationship is recovered? \textbf{RQ2:} What cryptographic primitives, system architectures, and operational procedures enable recovery? \textbf{RQ3:} What safety, liveness, privacy, usability, and trust tradeoffs are reported? \textbf{RQ4:} What open challenges remain for wallets, smart accounts, custody systems, decentralized identity, secure hardware, and related financial infrastructure?

\subsection{Discovery Corpus}
\label{subsec:discovery_corpus}
We searched Scopus, IEEE Xplore, the ACM Digital Library, and ScienceDirect for peer-reviewed English-language computer-science papers on cryptographic key backup, recovery, and restoration. The search was run and exported on 14 October 2025. The logical query was:

\begin{quote}
\small
(``key'' OR ``private key'' OR ``cryptographic key'') AND (``recovery'' OR ``backup'') AND (``public key infrastructure'' OR blockchain OR cryptocurrency OR wallet OR ``self-sovereign identity'' OR ``decentralized identity'' OR ``identity management'' OR ``digital identity'').
\end{quote}

The same logical query was adapted only for database field syntax and export format. Table~\ref{tab:database-search} reports the database-specific search surface and counts. The SLR log records 13,637 initial database-interface hits before database-side filters; the local export audit contains 4,322 BibTeX records and 4,144 normalized records. The round-0 screening log contains 4,314 records after basic filters and 4,044 after duplicate removal.

\begin{table}
  \centering
  \scriptsize
  \caption{Database search audit. The table separates library exports from normalized screening input so reviewers can trace the corpus before snowballing and SoK-stage filtering.}
  \label{tab:database-search}
  \begin{tabular}{@{}p{0.20\linewidth}rrp{0.42\linewidth}@{}}
    \hline
    Database & Exported & Normalized & Syntax and filter note \\
    \hline
    Scopus & 218 & 129 & \texttt{TITLE-ABS-KEY} logical query; Computer Science, English, article/conference filters \\
    IEEE Xplore & 462 & 434 & All-metadata logical query; English conference/journal records exported in BibTeX \\
    ACM Digital Library & 1,542 & 1,506 & Full-record or metadata logical query; ACM BibTeX batch exports \\
    ScienceDirect & 2,100 & 2,075 & Title/abstract/keywords logical query; ScienceDirect BibTeX batch exports \\
    \hline
  \end{tabular}
\end{table}

The initial title-and-abstract screening used a Large Language Model through an API as a decision-support tool: \texttt{gpt-5} for title/abstract screening and \texttt{gpt-5-mini} for basic filtering of snowballing references. Each prompt included the candidate title and abstract, inclusion and exclusion criteria, and a request for an accept/reject verdict plus justification. Outputs were recorded as status and comments fields. Both LLM-assisted inclusions and exclusions were manually verified, and accepted records were reconciled into the round-level BibTeX files before quality assessment. 
The model scaled screening; it did not make final inclusion decisions.

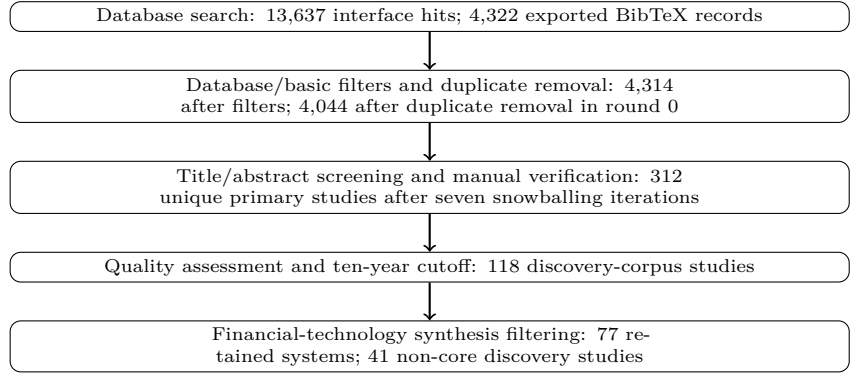
\begin{figure}[h]
  \centering
  \begin{tikzpicture}[
    node distance=0.50cm,
    box/.style={draw, rounded corners, align=center, text width=0.78\textwidth, inner sep=2.5pt, font=\scriptsize},
    arrow/.style={->, thick}
  ]
    \node[box] (hits) {Database search: 13,637 interface hits; 4,322 exported BibTeX records};
    \node[box, below=of hits] (filters) {Database/basic filters and duplicate removal: 4,314 after filters; 4,044 after duplicate removal in round 0};
    \node[box, below=of filters] (screening) {Title/abstract screening and manual verification: 312 unique primary studies after seven snowballing iterations};
    \node[box, below=of screening] (qa) {Quality assessment and ten-year cutoff: 118 discovery-corpus studies};
    \node[box, below=of qa] (sok) {Financial-technology synthesis filtering: 77 retained systems; 41 non-core discovery studies};
    \draw[arrow] (hits) -- (filters);
    \draw[arrow] (filters) -- (screening);
    \draw[arrow] (screening) -- (qa);
    \draw[arrow] (qa) -- (sok);
  \end{tikzpicture}
  \caption{Auditable corpus-construction flow. The key transition is from broad SLR discovery to a 77-system synthesis corpus used for taxonomy, matrix counts, and findings.}
  \label{fig:methodology-flow}
\end{figure}

Backward and forward snowballing were applied until convergence: a final iteration produced no new snowballing references and left the retained primary-study count unchanged at 312. Table~\ref{tab:snowballing-rounds} reports the round-level counts.

\begin{table}
  \centering
  \scriptsize
  \caption{Screening and snowballing counts. Round 7 provides the stopping rule. RA = research articles or candidate references; BF = after basic filters; RD = after duplicate removal; IE = cumulative retained records after inclusion/exclusion screening; BS = new backward/forward snowballing references.}
  \label{tab:snowballing-rounds}
  \begin{tabular}{@{}lrrrrr@{}}
    \hline
    Round & RA & BF & RD & IE & BS \\
    \hline
    0 & 13,637 & 4,314 & 4,044 & 129 & 1,820 \\
    1 & 1,949 & 1,147 & 817 & 230 & 1,390 \\
    2 & 1,620 & 849 & 623 & 281 & 668 \\
    3 & 950 & 594 & 478 & 303 & 208 \\
    4 & 511 & 401 & 360 & 308 & 67 \\
    5 & 375 & 331 & 321 & 311 & 21 \\
    6 & 332 & 321 & 315 & 312 & 5 \\
    7 & 317 & 314 & 313 & 312 & 0 \\
    \hline
  \end{tabular}
\end{table}

Table~\ref{tab:exclusion-reasons} summarizes grouped exclusion reasons. Title/abstract reasons were rule-coded from the manually verified decision comments. The final SoK-stage exclusions were coded from the full-text extraction rows.

\begin{table}
  \centering
  \scriptsize
  \caption{Grouped exclusion reasons. The table distinguishes title/abstract screening exclusions from full-text SoK-stage exclusions, showing why records left the evidence base.}
  \label{tab:exclusion-reasons}
  \begin{tabular}{@{}p{0.39\linewidth}rp{0.33\linewidth}r@{}}
    \hline
    Title/abstract exclusion reason & Count & SoK-stage exclusion reason & Count \\
    \hline
    Cryptographic security or authentication, but no recovery & 2,119 & Transferable non-financial background only & 28 \\
    Adjacent domain, but no recovery mechanism & 1,866 & Financial-adjacent but not core comparable recovery & 9 \\
    Review or non-primary literature & 434 & Requirements or background only & 4 \\
    Unrelated application or out of scope & 359 & & \\
    Key generation/protection, not recovery of an existing key & 279 & & \\
    No cryptographic private-key or ownership context & 108 & & \\
    Insufficient or missing abstract & 15 & & \\
    \hline
  \end{tabular}
\end{table}

\subsection{Quality Assessment and Discovery Corpus}
\label{subsec:quality_assessment}
After snowballing convergence, 312 primary studies were retained for quality assessment. Each study was scored using five criteria: evidence and validation; cryptographic foundations and techniques; security analysis and limitations; proposed solution or contribution; and open problems or future directions. Each criterion was scored as yes (1.0), partial (0.5), or no (0.0).

Applying the QA threshold retained 194 studies with score at least 4.0. Of these, 76 were pre-2016 foundational papers and are used as background where appropriate, while 118 had score at least 4.0 and publication year at least 2016. Those 118 studies form the discovery corpus used for full-text extraction and SoK narrowing.

\subsection{Financial-Technology Synthesis Corpus}
\label{subsec:aft_synthesis_corpus}
The 118-paper discovery corpus was intentionally broader than the final SoK needed to be. It included direct wallet and custody papers, but also adjacent work on generic key management, authentication, identity, hardware, cloud backup, and data protection. For the final synthesis, we therefore performed a second filtering step after full-text extraction.

Each extracted paper was coded for recovered object, secret/control recovery axis, mechanism family, enrollment and storage, recovery trigger, failure-event class, authorization, trust placement, adversary model, post-recovery state, financial-technology context, validation evidence, deployment status, privacy, usability, post-quantum relevance, and limitations. We retained papers that described a concrete recovery, backup, restoration, transfer, re-issuance, or extraction mechanism; were directly financial or clearly transferable to financial control; and contained enough detail to compare recovery semantics and tradeoffs. We excluded papers where recovery was peripheral, generic without a financial-technology transfer, or not comparable along the SoK dimensions.

This filtering reduced the corpus from 118 discovery papers to 77 synthesis papers. The excluded 41 papers were not treated as irrelevant; rather, they informed boundary-setting and background but were not used as core evidence for the taxonomy, counts, and cross-cutting findings. Figure~\ref{fig:methodology-flow} summarizes the full flow, while Table~\ref{tab:corpus-construction} summarizes the final two-stage construction.

\begin{table}
  \centering
  \small
  \caption{Two-stage corpus construction. Reviewers should read the 118 papers as the discovery corpus and the 77 retained papers as the evidence base for the SoK's comparative claims.}
  \label{tab:corpus-construction}
  \begin{tabular}{p{0.24\linewidth}p{0.48\linewidth}p{0.18\linewidth}}
    \hline
    Stage & Role in the paper & Result \\
    \hline
    SLR discovery & Identify peer-reviewed work on cryptographic key backup, recovery, and restoration through database search, manual verification, and snowballing. & 118 papers \\
    Financial-technology extraction & Retain papers with concrete recovery semantics and direct or transferable relevance to wallets, custody, smart accounts, decentralized identity, secure hardware, or financial infrastructure. & 77 papers \\
    \hline
  \end{tabular}
\end{table}

\section{Corpus Overview}
\label{sec:corpus-overview}

The 77-paper synthesis corpus is the evidence base for the taxonomy, mechanism comparison, and cross-cutting findings. It is narrower than the 118-paper discovery corpus because the goal is to systematize mechanisms that restore cryptographic or financial control, not to catalog every paper that mentions key recovery. The retained studies are concentrated in self-custody and blockchain wallet settings, with additional evidence from institutional custody, permissioned blockchains, smart-contract accounts, decentralized identity, secure hardware, and forensic recovery. Categories are non-exclusive.

\begin{table}[!t]
  \centering
  \small
  \caption{Non-exclusive corpus facets. Counts can exceed 77 within a facet when a study belongs to multiple categories.}
  \label{tab:corpus-facets}
  \begin{tabular}{p{0.22\linewidth}p{0.64\linewidth}}
    \hline
    Facet & Largest extracted groups \\
    \hline
    Financial context & Self-custody wallets (57), payment or permissioned blockchains (14), transferable non-financial patterns (15), institutional custody (11), DeFi or smart-contract applications (10), smart-contract accounts (9), decentralized identity for finance (8), secure hardware for financial infrastructure (8). \\
    Recovered object & Original private keys (34), signing authority (20), account control (19), seed or root secrets (15), asset ownership or transfer paths (13), key shares (10), recovery keys (5), credentials (3), forensic artifacts (3). \\
    Validation evidence & Prototypes (56), benchmarks (50), formal proofs (28), simulations (17), security games (11), case studies (9), gas-cost measurements (7), user studies (3), no evaluation in the extracted evidence (6), operational deployment evidence (1). \\
    \hline
  \end{tabular}
\end{table}

The venue and validation mix affects how the counts should be read. The corpus is recent-heavy: 7 of the 77 synthesis rows are from 2016--2018, while the remaining rows are from 2019 or later, with 43 conference or symposium papers, 32 journal articles, and 2 workshop or CEUR papers. This makes the evidence strongest for mechanism comparison, protocol feasibility, and engineering validation. It is weaker for deployment prevalence, recovery under stress, and operational or user outcomes, which is why we report user-study and deployment evidence separately instead of treating prototype, proof, benchmark, and deployment rows as equivalent. Table~\ref{tab:venue-confidence} reports the confidence tiers used to interpret the venue and validation mix.

\begin{table}[!b]
  \centering
  \scriptsize
  \caption{Venue and evidence confidence tiers used to interpret corpus counts. The tiers affect confidence in claims drawn from the counts, not whether a retained row is counted in the matrix.}
  \label{tab:venue-confidence}
  \begin{tabular}{@{}p{0.25\linewidth}rp{0.52\linewidth}@{}}
    \hline
    Confidence tier & Rows & Count interpretation \\
    \hline
    Higher for protocol soundness & 28 & Strongest for cryptographic claims, safety arguments, and mechanism feasibility; weaker for deployed prevalence or user behavior. \\
    Higher for external validity & 5 & Stronger support for user, forensic, or operational observations, but still narrow relative to the breadth of wallet and custody practice. \\
    Medium engineering evidence & 40 & Useful for comparing implemented prototypes, benchmarks, and design tradeoffs; not enough to infer production reliability. \\
    Lower or contextual evidence & 4 & Included for requirements, boundary-setting, or conceptual relevance; used cautiously and not as strong evidence for mechanisms. \\
    \hline
  \end{tabular}
\end{table}

The master matrix gives each retained system one auditable row covering the comparison dimensions, evidence fields, limitations, and SoK relevance. Table~\ref{tab:master-matrix-summary} compresses that matrix by the paper's primary organizing axis. The \href{https://github.com/franciscobecerra97/Cryptographic-Key-Recovery-for-Financial-Technologies/tree/aft-2026-artifacts} {reproducibility package} provides the complete machine-readable matrix, its codebook, summary tables, and audit artifacts.

\begin{table}[h]
  \centering
  \scriptsize
  \caption{Compressed view of the 77-system master matrix by recovery axis.}
  \label{tab:master-matrix-summary}
  \begin{tabular}{@{}p{0.13\linewidth}p{0.05\linewidth}p{0.23\linewidth}p{0.21\linewidth}p{0.27\linewidth}@{}}
    \hline
    Recovery axis & Rows & Dominant mechanisms & Typical post-recovery state & Studies \\
    \hline
    Secret-restoring & 38 & Threshold reconstruction, password/factor-protected recovery, biometric or fuzzy regeneration, TEE/HSM-assisted recovery & Same or equivalent secret available; occasional share refresh or credential reset & \cite{2,6,11,16,24,25,28,29,30,31,34,36,38,41,43,45,46,49,55,57,62,66,72,73,74,79,81,83,84,89,90,93,95,98,99,107,112,113} \\
    Hybrid secret/control & 16 & Deterministic regeneration, threshold continuity, institutional backup, social or guardian recovery & Secret material is restored while signing authority, custody participation, or account control may also change & \cite{19,27,42,53,56,68,80,82,85,94,97,101,106,111,117,118} \\
    Control-restoring & 16 & Smart-contract control transfer, fallback or ownership proof, passkey/backup authenticator recovery, pre-signed asset evacuation & New owner, credential, signing path, or asset-control relation replaces the failed path & \cite{8,10,18,23,37,40,54,59,63,75,86,87,96,102,108,114} \\
    Forensic/ extractive & 5 & Forensic artifact recovery and cryptanalytic key extraction & Artifact, credential, or key is extracted from device, backup, memory, nonce failure, or investigation evidence & \cite{14,44,71,76,103} \\
    Framework/ requirements & 2 & Requirements analysis and composability modeling & No single post-recovery state; evaluates recovery obligations or composed mechanisms & \cite{48,109} \\
    \hline
  \end{tabular}
\end{table}

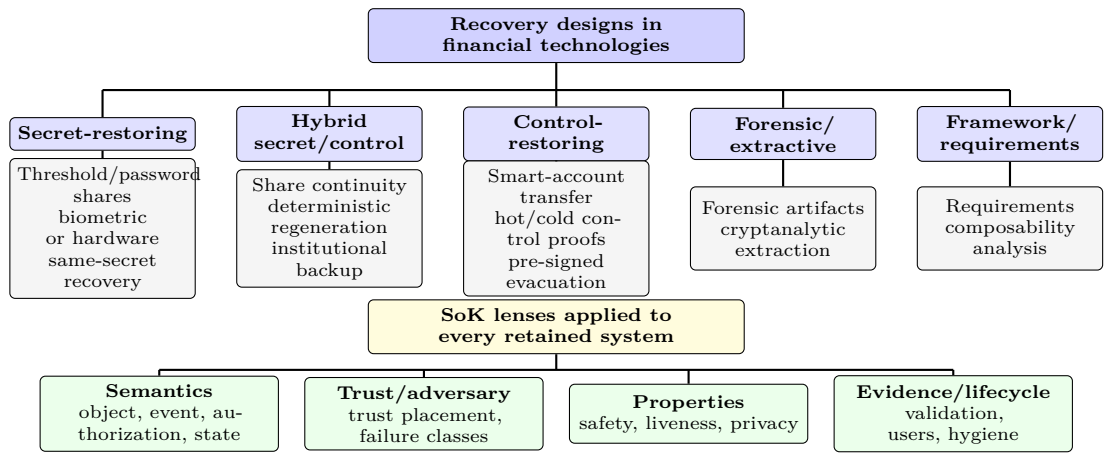
\begin{figure}[t]
  \centering
  \scriptsize
  \begin{tikzpicture}[
    box/.style={draw, rounded corners=2pt, align=center, inner sep=3pt, font=\scriptsize},
    root/.style={box, fill=blue!18, text width=0.34\linewidth, font=\bfseries\scriptsize},
    axis/.style={box, fill=blue!12, text width=0.16\linewidth, font=\bfseries\scriptsize},
    mech/.style={box, fill=gray!8, text width=0.16\linewidth, minimum height=1.10cm},
    lens/.style={box, fill=green!8, text width=0.21\linewidth, minimum height=0.78cm},
    lensroot/.style={box, fill=yellow!16, text width=0.34\linewidth, font=\bfseries\scriptsize},
    conn/.style={draw, thick}
  ]
    \node[root] (root) at (0,0) {Recovery designs in financial technologies};

    \node[axis] (secret) at (-6.0,-1.30) {Secret-restoring};
    \node[axis] (hybrid) at (-3.0,-1.30) {Hybrid secret/control};
    \node[axis] (control) at (0,-1.30) {Control-restoring};
    \node[axis] (forensic) at (3.0,-1.30) {Forensic/ extractive};
    \node[axis] (framework) at (6.0,-1.30) {Framework/ requirements};

    \draw[conn] (root.south) -- (0,-0.72);
    \draw[conn] (-6.0,-0.72) -- (6.0,-0.72);
    \foreach \x/\n in {-6.0/secret,-3.0/hybrid,0/control,3.0/forensic,6.0/framework}
      \draw[conn] (\x,-0.72) -- (\n.north);

    \node[mech] (secretm) at (-6.0,-2.55) {Threshold/password shares\\biometric or hardware\\same-secret recovery};
    \node[mech] (hybridm) at (-3.0,-2.55) {Share continuity\\deterministic regeneration\\institutional backup};
    \node[mech] (controlm) at (0,-2.55) {Smart-account transfer\\hot/cold control proofs\\pre-signed evacuation};
    \node[mech] (forensicm) at (3.0,-2.55) {Forensic artifacts\\cryptanalytic extraction};
    \node[mech] (frameworkm) at (6.0,-2.55) {Requirements\\composability analysis};

    \foreach \a/\m in {secret/secretm,hybrid/hybridm,control/controlm,forensic/forensicm,framework/frameworkm}
      \draw[conn] (\a.south) -- (\m.north);

    \node[lensroot] (lroot) at (0,-3.85) {SoK lenses applied to every retained system};
    \node[lens] (semantics) at (-5.25,-5.00) {\textbf{Semantics}\\object, event, authorization, state};
    \node[lens] (trust) at (-1.75,-5.00) {\textbf{Trust/adversary}\\trust placement, failure classes};
    \node[lens] (properties) at (1.75,-5.00) {\textbf{Properties}\\safety, liveness, privacy};
    \node[lens] (evidence) at (5.25,-5.00) {\textbf{Evidence/lifecycle}\\validation, users, hygiene};

    \draw[conn] (lroot.south) -- (0,-4.42);
    \draw[conn] (-5.25,-4.42) -- (5.25,-4.42);
    \foreach \x/\n in {-5.25/semantics,-1.75/trust,1.75/properties,5.25/evidence}
      \draw[conn] (\x,-4.42) -- (\n.north);
  \end{tikzpicture}
  \caption{Roadmap of the SoK taxonomy and lenses. The top tree groups recovery designs by the promise they make; the bottom lenses are applied to every retained system in the matrix and used in the tradeoff analysis.}
  \label{fig:sok-taxonomy-lenses}
\end{figure}

\section{Recovery Semantics and Evaluation Vocabulary}
\label{sec:recovery-semantics-vocabulary}

Recovery mechanisms in the corpus do not all recover the same thing: some reconstruct a private key or seed, some repair a threshold role, some rotate account control, and some extract artifacts after an incident. We therefore compare each design using four recovery-semantics dimensions: the \emph{recovered object}, the \emph{failure event}, the \emph{authorization path}, and the \emph{post-recovery state}. These dimensions describe what is restored, why recovery is invoked, what evidence or actors authorize it, and what control relation remains afterward. The adversary classes and recovery properties at the end of this section are supporting evaluation lenses, not additional parts of the four-part framework.

\subsection{Recovered object}
Recovery may reconstruct an original private key or seed, restore a share or signing role, restore account control or asset ownership, or extract forensic artifacts. Mechanisms include password-protected shares, threshold reconstruction, deterministic regeneration, biometric or physical regeneration, hardware-assisted retrieval, contracts, fallback keys, backup authenticators, pre-signed transactions, and ownership proofs. Table~\ref{tab:master-matrix-summary} lists retained studies by recovery axis, Appendix~\ref{app:master-matrix} gives the full row-level coding, and Table~\ref{tab:mechanism-taxonomy} gives representative mechanism examples.

\subsection{Failure event}
The corpus covers several failure modes: loss of a local key, seed, device, mnemonic, or share; unavailability of threshold participants or custody roles; loss or compromise of smart-account owners; failed identity-wallet migration or backup restoration; and post-incident forensic recovery. Table~\ref{tab:family-adversary} summarizes how these failure modes map to adversary classes. A recurring limitation is that many designs focus on loss but give less detail for compromise, coercion, malware, or malicious recovery helpers.

\subsection{Authorization path}
Recovery is authorized by different evidence of legitimacy. Depending on the family, authorization may require enough shares or servers, a memorized or factor-derived credential, retained root material, biometric helper data, guardian approval, contract timeouts and challenges, pre-authorized transactions, organizational procedures, hardware controls, or custody roles. The authorization basis matters because it determines who can approve recovery, who can block it, and what evidence remains after the event.

\subsection{Post-recovery state}
The post-recovery state is often under-specified. Some protocols return the same secret and continue with unchanged keys \cite{2,11}; some repair or refresh shares while preserving threshold signing authority \cite{27,94}; some restore access by rotating credentials or moving control to a new key \cite{8,87}; and some transfer assets away from a failing account \cite{75,114}. For financial technologies, this distinction is material: a user, exchange, or smart account that rotates ownership after recovery has a different audit and dispute problem from one that reconstructs the old private key.

\subsection{Supporting adversary vocabulary}
We also normalize the threat vocabulary used across papers, because terms such as ``lost key,'' ``recovery,'' and ``compromise'' otherwise mix different financial risks. We distinguish eight recurring classes. \emph{Loss} means the legitimate party no longer has usable access to a key, seed, credential, device, share, passkey, or signer. \emph{Compromise or theft} means that the material may be copied, stolen, malware-exposed, or attacker-controlled. \emph{Coercion} means that a user, guardian, provider operator, or institutional signer is pressured to approve recovery or disclose recovery material. A \emph{malicious helper} is a guardian, shareholder, trustee, server, or custody participant who deviates from the recovery protocol alone or in collusion. \emph{Server or provider abuse} covers wallet providers, cloud backup services, identity verifiers, recovery providers, HSM operators, and infrastructure services that misuse their role or metadata. An \emph{institutional insider} is an organizational actor such as an administrator, treasurer, support agent, compliance role, or recovery officer who misuses delegated recovery authority. \emph{Unavailability} means that a user, helper, server, blockchain, signer, network, device, or institution cannot participate when needed. A \emph{forensic adversary} attempts to extract wallet artifacts, keys, backups, traces, or evidentiary control material from devices or services.

\subsection{Recovery-specific properties}
Finally, we evaluate each design through three recovery-specific properties. \emph{Safety} means unauthorized parties cannot trigger recovery, reconstruct a secret, rotate control, evacuate assets, or exploit recovery metadata to steal financial control. \emph{Liveness} means a legitimate claimant can complete recovery after loss, participant churn, device replacement, institutional outage, or disaster without relying on a single fragile actor. \emph{Privacy} means the recovery process does not unnecessarily reveal helper relationships, provider choices, account linkage, recovery attempts, identity-proofing material, device replacement, or forensic traces.

\section{Taxonomy of Key Recovery Mechanisms}
\label{sec:taxonomy}

Table~\ref{tab:mechanism-taxonomy} is axis-first: mechanism families are nested under the recovery semantics they primarily realize. This avoids grouping systems once by cryptographic primitive and again by design family. The primary recovery axis is mutually exclusive and assigned by the final recovery promise and post-recovery state recorded in Appendix~\ref{app:master-matrix}. Mechanism families remain non-exclusive because some systems combine mechanisms, such as threshold reconstruction inside a hardware wallet or backup-authenticator recovery inside a smart account. Citation columns are representative; exhaustive row-level coverage appears in Appendix~\ref{app:master-matrix}.

\begin{table}[!h]
  \centering
  \scriptsize
  \caption{Axis-first taxonomy and design comparison. Mechanisms that reconstruct a secret, rotate control, move assets, or extract evidence make different recovery promises.}
  \label{tab:mechanism-taxonomy}
  \begin{tabular}{@{}p{0.12\linewidth}p{0.19\linewidth}p{0.24\linewidth}p{0.23\linewidth}p{0.13\linewidth}@{}}
    \hline
    Axis & Mechanism family & What recovery restores & Enrollment or storage pattern & Examples \\
    \hline
    Secret-restoring & Threshold, password-protected, and factor-assisted reconstruction & Same key, seed, mnemonic fragment, or share & Distributed shares, password/factor material, protected server shares, backup fragments & \cite{2,6,11,16,24,49,79,89,90,98} \\
    Secret-restoring & Memory-based, biometric, fuzzy, and hardware-assisted regeneration & Same or equivalent wallet-control secret & Memory artifact, helper data, biometric source, secure element, HSM, or TEE & \cite{30,34,55,66,81,93,95,99} \\
    Hybrid secret/control & Share continuity, social/delegated recovery, biometric helper-key regeneration, and institutional disaster recovery & Secret material plus signing role, custody role, account access, or wallet-governance authority & Committees, guardians, biometric helper keys, organizational media, custody procedures, or DID backup stores & \cite{19,27,53,82,94,101,106,111,117} \\
    Control-restoring & Deterministic hot/cold signing-authority recovery & Future signing authority after hot-key compromise, without reconstructing the compromised hot key & Cold-wallet seed or master state, derivation path, and synchronized wallet state & \cite{23} \\
    Control-restoring & Smart-contract transfer, fallback signing, passkey recovery, credential re-issuance, and ownership proof & Account owner, signer, backup credential, ownership proof, or control path & On-chain state, fallback keys, backup authenticators, challenge windows, re-issuance, or proofs & \cite{8,10,37,40,59,86,87,96,102} \\
    Control-restoring & Asset evacuation and pre-authorized transfer & Value or asset-control relation moved away from a failed path & Pre-signed transactions, emergency controls, provider state, or encrypted transfer material & \cite{75,114} \\
    Forensic/ extractive & Forensic artifact recovery and cryptanalytic extraction & Wallet artifacts, credentials, keys, or evidentiary traces & Device access, backups, memory, nonce failures, investigation workflow, or seized evidence & \cite{14,44,71,76,103} \\
    Framework/ requirements & Requirements and composability analysis & No single recovered object; compares obligations or composed mechanisms & Requirements model, key-person analysis, or formal composition model & \cite{48,109} \\
    \hline
  \end{tabular}
\end{table}

The primary split is between \emph{secret-restoring} and \emph{control-restoring} recovery. Secret-restoring mechanisms preserve the old control relation by reconstructing or regenerating the same key, seed, or share. Control-restoring mechanisms intentionally create a new control relation, for example by moving an account owner, registering a backup authenticator, executing a pre-signed transaction, or proving ownership to migrate assets. Hybrid systems combine both promises, such as restoring a share while also preserving institutional signing authority or social-account control. Control-restoring designs are not weaker merely because they do not reconstruct the original secret; they make a different promise and need different post-recovery checks. Figure~\ref{fig:sok-taxonomy-lenses} summarizes the taxonomy spine and the lenses applied to each retained system.

The taxonomy also separates \emph{cryptographic trust distribution} from \emph{operational trust distribution}. Threshold systems distribute cryptographic material but still need participant governance, refresh, authentication, and availability. Hardware systems reduce software exposure but concentrate trust in devices, attestation, or recovery artifacts. Social recovery and smart contracts reduce single-custodian dependence but add guardian, metadata, liveness, and irreversibility risks.

\section{Mechanism-Family Tradeoff Analysis}
\label{sec:mechanism-family-tradeoffs}

Using the matrix in Appendix~\ref{app:master-matrix} and the axis-first taxonomy, we compare safety and liveness surfaces, post-recovery hygiene and metadata privacy, and validation evidence. Table~\ref{tab:mechanism-taxonomy} gives the axis-level mechanism comparison, Table~\ref{tab:family-adversary} adds the adversary lens, and Table~\ref{tab:hygiene-privacy} summarizes the obligations that follow after recovery. The family tables are cross-cutting lenses rather than primary-axis reclassifications; each cited system keeps the canonical axis assigned in Appendix~\ref{app:master-matrix}.

\begin{table}[!h]
  \centering
  \scriptsize
  \caption{Family-by-adversary comparison. Actors who improve liveness can also become collusion, coercion, provider, insider, or metadata surfaces.}
  \label{tab:family-adversary}
  \begin{tabular}{@{}p{0.14\linewidth}p{0.16\linewidth}p{0.19\linewidth}p{0.18\linewidth}p{0.20\linewidth}p{0.08\linewidth}@{}}
    \hline
    Family & Loss / unavailability & Compromise / theft & Coercion / malicious helpers & Provider, insider, metadata, or forensic exposure & Examples \\
    \hline
    Threshold and password-protected reconstruction & Lost keys, shares, or passwords when enough shares/servers remain live & Exposed shares, weak passwords, or stale fragments can reconstruct & Collusion or coercion crossing the threshold & Server contact, helper membership, and recovery attempts may leak; hosted shares add provider trust & \cite{2,6,11,16,24,62,79,89,90,94} \\
    Deterministic, biometric, and hardware recovery & Device, seed, mnemonic, signer, or factor loss when source material remains available & Restored seeds, helper keys, or cold-wallet authority may preserve attacker access unless old paths are quarantined & Biometrics, memories, devices, or support flows can be coerced & Vendors, attestation, HSMs, helper data, and identity checks become trust/privacy points & \cite{23,30,34,55,66,81,93,95,99,111,112} \\
    Social, guardian, and smart-account transfer & Lost owners, passkeys, or signers when guardians, fees, and windows are available & Compromised guardians, fallback signers, modules, or passkeys can rotate control & Vote buying, social engineering, recoverer abuse, and guardian collusion & On-chain proposals, guardian sets, and signer rotations can reveal distress and relationships & \cite{8,10,37,40,53,59,75,82,86,87,96,102,108,109,114} \\
    Institutional custody and disaster recovery & Signer churn, lost institutional devices, outages, and disasters & Compromised operator devices, HSM credentials, backups, or policy engines can move funds & Collusion or coercion of officers can satisfy dual-control policies & Administrators, recovery officers, compliance roles, logs, and drills expose insider/metadata surfaces & \cite{18,48,85,101,117} \\
    Forensic and extractive recovery & Inaccessible devices, backups, memory, nonce failures, or investigation evidence & The same extraction path can be abused by attackers or overbroad investigators & Coercion may compel device access, backup disclosure, or support escalation & Forensic traces are often the recovered object; chain-of-custody and privacy risks are intrinsic & \cite{14,44,71,76,103} \\
    \hline
  \end{tabular}
\end{table}

\begin{table}[h]
  \centering
  \scriptsize
  \caption{Post-recovery hygiene and metadata/privacy obligations by mechanism family.}
  \label{tab:hygiene-privacy}
  \begin{tabular}{@{}p{0.17\linewidth}p{0.24\linewidth}p{0.25\linewidth}p{0.22\linewidth}p{0.07\linewidth}@{}}
    \hline
    Family & Post-recovery hygiene & Metadata/privacy exposure & Common gap & Examples \\
    \hline
    Threshold and password-protected reconstruction & Rotate/rekey if compromise is possible; revoke stale fragments; refresh shares; audit helpers/servers & Helper linkage, server contact, timing, and repeated use can reveal distress or linkage & Old-secret reconstruction often lacks rotation, revocation, or dispute handling & \cite{2,6,11,16,24,62,79,89,90,94} \\
    Deterministic, biometric, and hardware recovery & Revoke backups/helper data; reset authenticators; quarantine lost devices; migrate funds or rotate control if old material may still authorize value & Device replacement, helper data, vendor contact, identity checks, and account linkage expose context & Many designs lack a decision rule for continued use of restored material or retirement of old authority & \cite{23,30,34,55,66,81,93,95,99,111,112} \\
    Social, guardian, and smart-account transfer & Revoke old owners/passkeys; refresh guardians; review delays/cancellation; audit final owner state & Guardian relationships, proposals, signer rotations, challenge windows, and module state reveal attempts & Guardian reset, cancellation, rollback, and dispute handling are uneven & \cite{8,10,37,40,53,59,75,82,86,87,96,102,108,109,114} \\
    Institutional custody and disaster recovery & Rotate signer/API/HSM credentials and policy roles; preserve logs; quarantine affected devices & Approver graphs, provider contacts, audits, drills, KYC, insurance, and legal holds reveal operations & Governance, notification, and disputed-instruction workflows are less uniform & \cite{18,48,85,101,117} \\
    Forensic and extractive recovery & Move funds or rotate keys; revoke recovered credentials; maintain chain-of-custody; quarantine devices/secrets & Device images, backups, artifacts, clusters, investigator actions, and asset movement are metadata-rich & Legitimate recovery and adversarial extraction share tooling, so authority boundaries matter & \cite{14,44,71,76,103} \\
    \hline
  \end{tabular}
\end{table}

The first comparison axis is \emph{safety}: who can cause unauthorized recovery? Threshold systems bound exposure by shares or servers but depend on enrollment, participant independence, and refresh; smart-account and guardian systems avoid custodial key storage but add vote-buying, social-engineering, coercion, and challenge-window risks; hardware designs isolate secrets but depend on device, attestation, backup, or HSM assumptions. The second axis is \emph{liveness}: threshold, contract, biometric, memory-based, and institutional systems can all fail when helpers, fees, inputs, terminals, or policies are unavailable. The third axis is \emph{evidence}: the corpus validates feasibility and performance more often than recovery under stressful, high-value conditions.

\section{Generalized Construction Model}
\label{sec:construction-model}

Across the taxonomy, recovery designs can be described by a common construction model. It is not a protocol template; it is a checklist of design obligations that recur across wallets, custody systems, identity wallets, and smart accounts.

\begin{enumerate}
    \item \textbf{Control material.} The system identifies what is at risk: a private key, seed, key share, recovery key, credential, signing role, account owner, or asset-control relation.
    \item \textbf{Enrollment.} The system creates recovery material before failure, such as secret shares \cite{6,11,49}, masked password-protected shares \cite{2,16,41}, deterministic seeds \cite{23,42,93}, biometric helper data \cite{81,95,111}, hardware-wrapped secrets \cite{30,55,99}, guardians \cite{19,37,82}, backup authenticators \cite{86,102,112}, or pre-signed transactions \cite{75,114}.
    \item \textbf{Storage and governance.} Recovery material is placed with the user, helpers, servers, institutions, smart contracts, secure hardware, cloud providers, or forensic evidence stores. This step determines whether the design is non-custodial, threshold-custodial, institutionally governed, or provider-assisted.
    \item \textbf{Failure trigger.} Recovery begins after loss, unavailability, compromise, migration, death, device replacement, identity-wallet restoration, or investigation. The trigger may be user-initiated, contract-triggered, institutional, or forensic.
    \item \textbf{Authorization and checks.} The system verifies that the recovery claimant is legitimate. Authorization can be cryptographic, social, institutional, on-chain, biometric, hardware-backed, or evidentiary. Strong designs separate authorization from mere possession of backup data.
    \item \textbf{Recovery action.} The system reconstructs a secret, regenerates a derived key, repairs a share, reissues a credential, rotates ownership, transfers assets, or extracts an artifact.
    \item \textbf{Post-recovery hygiene.} The system should decide whether to rotate keys, invalidate stale shares, revoke old authenticators, refresh guardians, record an audit event, or migrate funds. Many retained studies do not fully specify this stage, which is a central gap for financial deployments.
\end{enumerate}

This construction model explains why apparently similar mechanisms differ in financial risk. A threshold wallet that reconstructs the old seed \cite{11,89,90} exposes the recovered secret after completion. A smart-account recovery protocol that rotates ownership \cite{8,10,87} avoids exposing the old key but must secure the transfer process. A pre-signed evacuation mechanism \cite{75,114} may preserve asset value while abandoning the old account. Appendix~\ref{app:seven-stage-walkthroughs} walks four representative systems through all seven stages to show where their post-recovery hygiene obligations diverge. These semantics should not be evaluated under a single ``key recovery'' label.

\section{Production Recovery Designs}
\label{sec:production-recovery}

The 77-paper synthesis corpus remains the peer-reviewed evidence base for this SoK. Deployed wallet and custody systems are nevertheless useful practice checks: they are not counted as corpus rows, but they show whether the taxonomy and construction model cover recovery semantics used by retail users, institutional customers, and smart-account developers. Table~\ref{tab:production-recovery} classifies representative systems using official documentation.

\begin{table}[!h]
  \centering
  \scriptsize
  \caption{Production and standards-based practice checks outside the 77-paper corpus.}
  \label{tab:production-recovery}
  \begin{tabular}{@{}p{0.16\linewidth}p{0.14\linewidth}p{0.18\linewidth}p{0.28\linewidth}p{0.15\linewidth}@{}}
    \hline
    Design & Evidence source & Recovery axis and mechanism & Recovery action & Main SoK implication \\
    \hline
    Ledger Recover & Vendor product docs & Secret-restoring; secure-element export, encrypted fragments, HSM-backed providers & Identity checks release fragments to reconstitute seed entropy inside a Ledger secure element \cite{ledgerRecover2026} & Restores seed-equivalent material; moves trust to identity proofing, firmware consent, providers, and HSMs \\
    MPC custody & Vendor developer and help docs & Hybrid secret/control; MPC share backup, signer reprovisioning, disaster recovery & Fireblocks and Coinbase Prime describe share/signer recovery through provider infrastructure, devices, passphrases, and policy \cite{fireblocksDrs2026,fireblocksNcwBackup2026,coinbasePrimeOnchain2026} & Threshold cryptography still needs governance for passphrases, devices, signers, and disaster drills \\
    Safe \{RecoveryHub\} and modules & Safe documentation & Control-restoring; delayed recoverer module & A Recoverer proposes owner or threshold changes; signers can cancel during the delay \cite{safeRecoveryHub2026,safeModules2026} & Smart-account recovery migrates control and exposes on-chain recovery metadata \\
    ERC-4337 / EIP-7702 & Standards and protocol docs & Control-restoring substrate; custom validation and EOA delegation & ERC-4337 supports custom account validation; EIP-7702 lets EOAs delegate to code using authorization tuples \cite{erc4337,eip7702,ethereumPectra7702} & Account abstraction expands recovery design space; EIP-7702 leaves the EOA key as a bypass risk \\
    WebAuthn/ passkeys & W3C, FIDO, and product docs & Control-restoring; backup credential or passkey signer replacement & WebAuthn exposes backup state; Coinbase Smart Wallet uses a recovery phrase to create a new signer and passkey \cite{webauthn3,fidoSyncedPasskeys2024,coinbaseSmartWalletRecovery2026} & Passkey recovery depends on credential-provider recovery, fallback policy, and signer cleanup \\
    Argent guardian recovery & Wallet support docs & Control-restoring; guardian social recovery with delay & Guardians approve recovery; Argent Vault recovery has majority approval and a 48-hour delay \cite{argentGuardianRecovery2025,argentAddGuardian2025} & Social recovery improves liveness but creates guardian availability, coercion, and relationship-metadata questions \\
    \hline
  \end{tabular}
\end{table}

\paragraph*{Representative production recovery workflows}
The seven-stage model separates production systems that all call themselves ``recovery.'' Ledger Recover restores seed-equivalent material through device-approved fragmentation, provider-held encrypted fragments, and identity verification. Safe\{RecoveryHub\} stores recoverer and delay state on-chain so a recoverer can propose owner or threshold changes unless current signers cancel. MPC custody restores a share or signer path across customer devices, provider infrastructure, passphrases, and policy. Passkey smart wallets add or replace a signer, after which stale passkeys and recovery phrases remain asset-control material.

\section{Cross-Cutting Findings for Financial Technologies}
\label{sec:cross-cutting-findings}

\subsection{Finding 1: Recovery semantics are heterogeneous}
Recovery may mean reconstructing a key, regenerating a seed, restoring a share, recovering a recovery key, rotating credentials, restoring account control, transferring assets, or extracting forensic artifacts. The largest recovered-object groups are original private keys (34 studies), signing authority (20), account control (19), seeds or root secrets (15), asset ownership or transfer paths (13), and key shares (10). Counts are non-exclusive, but the design implication is clear: systems that restore account control should not be evaluated as if they reconstruct the original key.

\subsection{Finding 2: Non-custodial recovery still moves trust somewhere}
In this corpus, recovery mechanisms reduce or redistribute trust rather than eliminating it. Password-protected systems trust server independence and rate limiting; threshold systems trust participant diversity and share availability; hardware systems trust devices, attestation, or HSM boundaries; social recovery trusts guardians; and smart-contract recovery trusts public rules and challenge windows. In the trust and adversary fields, collusion appears in 22 rows, malicious actors in 20, server terms in 20, provider terms in 8, and insider terms in 8. These are conservative text-coded signals rather than a threat prevalence measure, but Table~\ref{tab:family-adversary} makes the trust shift explicit. A useful recovery design should state where trust has moved.

\subsection{Finding 3: Liveness improvements often create abuse paths}
Adding helpers, backup authenticators, pre-signed transactions, cloud backups, or institutional procedures can prevent permanent loss while creating coercion, collusion, phishing, insider, or unilateral-transfer paths. Unavailability appears in 16 failure-event rows, while compromise or theft appears in 16 rows; the adversary text mentions collusion in 22 rows, malicious actors in 20, and coercion in 8. This tension appears across social recovery \cite{19,37,82}, smart-account transfer \cite{8,10,59,87}, institutional custody \cite{48,101}, and hosted or provider-assisted identity recovery \cite{85,102,106}. Because coercion is rarely formalized, the observed coercion count is best read as a lower bound.

\subsection{Finding 4: Post-recovery lifecycle is a common blind spot}
Many studies explain how to recover but not what happens next: whether to rotate keys, refresh shares, revoke helpers, clear backup authenticators, audit the event, quarantine a device, migrate funds, or handle disputes. The post-recovery-state coding is dominated by same-or-equivalent secret restoration (46 rows), followed by new control relations or credentials (17), forensic extraction (4), operational access restored or rekeyed (3), analysis-only rows (3), asset movement (2), and share or committee refresh (2). Table~\ref{tab:hygiene-privacy} shows that same-secret systems need rotation or migration decisions, guardian systems need owner and helper reset, institutional systems need audit and policy cleanup, and forensic systems need chain-of-custody and authority boundaries.

\subsection{Finding 5: User evidence lags behind protocol evidence}
Formal proofs, prototypes, and benchmarks are common, but user-study evidence is rare. The extraction records 56 prototypes, 50 benchmarks, 28 formal proofs, 17 simulations, 11 security-game analyses, 9 case studies, 7 gas-cost measurements, and 6 rows with no evaluation in the extracted evidence. Only three retained studies include user-study evidence \cite{6,66,108}, and one forensic tool reports operational deployment evidence \cite{103}. Because recovery is invoked under stress, device loss, death, institutional disruption, or suspected compromise, we treat usability claims without user data as claims, not as user evidence.

\subsection{Finding 6: Recovery metadata is underprotected}
Several designs protect key material while saying less about metadata: which guardians exist, which servers were contacted, whether a recovery attempt occurred, which device was replaced, which account is being migrated, or which institution is under stress. Privacy or private-state terminology appears in 51 privacy-discussion rows, but 52 rows also include limited, no-formal-analysis, not-stated, or out-of-scope caveats; metadata, contact, timing, log, linkage, trace, or surveillance terms appear in 33 rows. These counts reflect conservative coding of what papers report, not proof that every omitted discussion leaks data. The issue cuts across threshold servers \cite{2,62,89}, guardians \cite{19,82}, smart accounts \cite{10,59,87}, identity wallets \cite{94,102,106}, and forensic tools \cite{71,103}.

\section{Financial, Regulatory, and Human-Factors Implications}
\label{sec:financial-human-implications}

\paragraph*{Custody, compliance, and operational controls}
Recovery becomes a regulated financial-technology concern when a provider holds, administers, restores, or can influence access to crypto-assets. MiCA-style custody rules, FATF virtual-asset guidance, and DORA operational-resilience obligations bring custody policy, client authentication, records, incident recovery, and provider duties into the design surface \cite{mica2023,ebaMicaArticle75,fatfVaspGuidance2021,dora2022}. This paper is not a legal analysis; the design point is that each mechanism should state whether recovery is self-custodial, provider-assisted, sub-custodial, institutional, or only a user-held backup. Obligations change depending on who verifies identity, sees recovery attempts, controls logs, and can block or reverse a disputed recovery \cite{18,48,85,101,117}.

\paragraph*{Travel Rule metadata}
Recovery designs that transfer assets, restore hosted access, or involve a crypto-asset service provider can also trigger Travel Rule information flows. FATF treats the Travel Rule as the virtual-asset application of Recommendation 16, requiring originator and beneficiary information to accompany covered transfers, while the EU implements parallel requirements for transfers of funds and certain crypto-assets in Regulation 2023/1113 and EBA guidelines \cite{fatfVaspGuidance2021,fatfRec16Update2025,euTravelRule2023,ebaTravelRuleGuidelines2024}. For recovery, the design implication is that identity proofing, beneficiary changes, emergency transfers, provider logs, and self-hosted-address interactions may become compliance metadata. Systems should therefore separate recovery authorization from Travel Rule reporting, minimize unnecessary recovery metadata, and state who can disclose records during disputes or investigations.

\paragraph*{Economics and incentives}
Recovery is also an incentive system. Guardians, shareholders, providers, support agents, and institutions may be unpaid, compensated, bonded, insured, or penalized; attackers can exploit those incentives through bribery, coercion, fee griefing, or collusion. Rational secret-sharing rows model some behavior \cite{89,90}, composability work makes rational adversaries explicit \cite{109}, and pre-signed evacuation exposes fee drift and stale-UTXO issues \cite{114}. Designs should specify who pays recovery fees, who absorbs disputed losses, and whether insurance or reimbursement changes diligence.

\paragraph*{Inheritance and death or incapacity}
Death and incapacity are not ordinary key-loss events: authority may shift to an heir, executor, trustee, provider, court, or pre-authorized beneficiary. The corpus contains only a small inheritance thread--key-person loss \cite{48}, biometric/heir wallet governance \cite{111}, re-registration recovery \cite{113}, and pre-signed recovery for owner disappearance or death \cite{114}. Estate recovery therefore needs separate treatment for enrollment while alive, beneficiary changes, disputed claims, privacy, and post-transfer auditability.

\paragraph*{Post-quantum migration}
NIST finalized the first three post-quantum cryptography FIPS in 2024 and selected HQC for future standardization as a backup key-establishment algorithm in 2025 \cite{nistPqcFips2024,nistHqc2025}. Recovery material is migration-sensitive because wallet seeds, encrypted mnemonics, guardian secrets, recovery tokens, custody shares, and pre-signed emergency transactions may outlive active signing keys. Same-secret restoration can reintroduce classical ECDSA or Schnorr keys after migration, so recovery should bind restored material to algorithm/version metadata and require re-derivation, fund migration, or key rotation when old material controls quantum-vulnerable addresses. Control-restoring recovery must update every path, not only the active signer: stale EOAs, guardians, fallback keys, passkeys, authorizations, and pre-signed transactions can remain bypasses. Institutional and forensic recovery also need crypto-agile policies for shares, approvals, HSM rules, audit signatures, and evidentiary records. The corpus has early signals \cite{40,93,96,102,112}, but no complete migration playbook for deployed wallets or custody systems.

\begin{table}[!h]
  \centering
  \scriptsize
  \caption{User-study evidence in the retained corpus. Existing studies test tangible shares, memory cues, and multi-factor wallet recovery, but not high-stress, adversarial, estate, or broad-population recovery.}
  \label{tab:user-study-evidence}
  \begin{tabular}{@{}p{0.16\linewidth}p{0.24\linewidth}p{0.25\linewidth}p{0.27\linewidth}@{}}
    \hline
    Study & Participants and task & Main finding & Main limitation for recovery claims \\
    \hline
    Offline secret sharing \cite{6} & 24-person form-factor survey; 18 completed field-study participants distributed five physical shares to 90 confidants and later retrieved enough shares for a 3-of-5 threshold & Physical share distribution and retrieval was practical, but recovery coordination could take days or weeks; paper versus plastic key tags did not dominate success & Student sample, one threshold configuration, simulated cryptographic reconstruction, and limited compliance checking \\
    Reminisce \cite{66} & 51 participants/510 pictures for parameters, 105-person survey, and 20 wallet-experienced users who repeated picture-location recovery after one month & All 20 recovered successfully; 18 succeeded on the first attempt, with 1.10--8.29 minute recovery times and 2.56 minutes average & Repository security and authentication out of scope; fixed parameters; 20--40-year-old experienced wallet users; photo/location privacy leakage \\
    Multi-factor key-derivation wallet \cite{108} & 27 United States participants compared custodial, conventional non-custodial, and MFKDF-based Ethereum wallet prototypes across creation, login, and recovery tasks & The MFKDF wallet reported 37\% higher System Usability Scale (SUS) and 71\% faster task completion than the conventional non-custodial wallet, with less seed-phrase handling & Small US-only and not gender-balanced sample; prototype fidelity limits; accessibility and marginalized populations not evaluated \\
    \hline
  \end{tabular}
\end{table}

Together, these studies show that tangible shares, picture-location memory, and familiar multi-factor workflows can work in controlled settings, but not how users behave under theft, grief, coercion, provider lockout, legal dispute, or market stress. Open questions include ceremony comprehension, stress errors, guardian coordination, inheritance, phishing, support escalation, accessibility, and recovery-metadata privacy.

\section{Open Problems and Research Agenda}
\label{sec:research-agenda}

\begin{itemize}
    \item \textbf{Standard recovery semantics.} Wallets, custody systems, and identity wallets need explicit terminology for original-key reconstruction, seed regeneration, share repair, credential re-issuance, control transfer, and asset evacuation. Without this, designs with different promises are difficult to compare.
    \item \textbf{Custody, regulation, and incentives.} Recovery mechanisms should declare whether they create self-custody, provider-assisted custody, sub-custody, or institutional obligations, and should model guardian availability, provider fees, Travel Rule metadata, reimbursement, insurance, fee drift, and collusion \cite{89,90,109,114,mica2023,ebaMicaArticle75,fatfVaspGuidance2021,euTravelRule2023,dora2022}.
    \item \textbf{Inheritance and estate recovery.} Death and incapacity require different authorization semantics from ordinary loss, including beneficiary enrollment, revocation while alive, dispute handling, privacy, and auditability for heirs, executors, trustees, and providers \cite{48,111,113,114}.
    \item \textbf{Composable recovery after loss or compromise.} Smart-account recovery increasingly combines guardians, backup authenticators, fallback signatures, timeouts, ownership proofs, and pre-signed transactions \cite{10,40,86,96,114}. Research is needed on how these paths compose, and on recovery that rotates, revokes, audits, and disputes after compromise rather than only restoring after loss \cite{2,95,107}.
    \item \textbf{Privacy-preserving recovery coordination.} Recovery should hide helper relationships, server choices, attempts, and account linkage where possible, especially for threshold servers, guardians, identity wallets, and smart-contract metadata.
    \item \textbf{Human-centered recovery under stress.} Recovery ceremonies should be evaluated with users and operators under time pressure, grief, device loss, or fear of theft. Existing user evidence is useful but too sparse and controlled for the importance of the problem \cite{6,66,108}.
    \item \textbf{Operational threshold custody.} Threshold protocols need lifecycle support for lost shares, changing personnel, institutional disaster recovery, participant churn, and auditability \cite{27,94,101}.
    \item \textbf{Long-lived recovery material.} Wallet seeds, backup credentials, custody shares, and pre-signed recovery paths may survive for years. Post-quantum recovery is emerging \cite{93,102,nistPqcFips2024}, while forensic recovery raises privacy, chain-of-custody, and abuse questions \cite{44,71,76,103}.
\end{itemize}

\section{Ethics, Limitations, and Reproducibility}
\label{sec:ethics-limitations-reproducibility}

This SoK studies mechanisms that can restore or transfer financial control. The ethical risk is therefore not limited to confidentiality of keys. Recovery systems can enable theft, coercion, unauthorized transfer, insider abuse, account lockout, or exclusion from identity and financial services. For this reason, the taxonomy distinguishes what is recovered and who authorizes recovery; ambiguity in these two points can hide serious harms.

The review has methodological limitations. The search considered peer-reviewed journal and conference articles written in English and classified as computer science. The final synthesis retained 77 of the 118 discovery papers after full-text extraction and financial-technology-focused filtering. LLM assistance was used as a decision-support tool during screening, and all LLM-assisted decisions were manually verified. The resulting corpus is suitable for a focused SoK, but it does not exhaust gray literature, wallet documentation, standards drafts, or production incidents.

The extraction is also limited by the evidence reported in each paper. If a study did not state a deployment, user evaluation, privacy analysis, or post-recovery rotation procedure, we classify that evidence as absent rather than infer it. This conservative choice may understate properties implemented outside the paper, but it avoids giving recovery mechanisms credit for unstated guarantees.

For reproducibility, the paper reports the source databases, exact logical query, search/export date, database counts, de-duplication and screening counts, grouped exclusion reasons, snowballing rounds, QA threshold, retained-study count, and extracted comparison dimensions. The \href{https://github.com/franciscobecerra97/Cryptographic-Key-Recovery-for-Financial-Technologies/tree/aft-2026-artifacts} {reproducibility package} contains search-count summaries, screening-round summaries, the LLM screening protocol and prompts, exclusion-reason summaries, QA counts, extracted rows, the matrix codebook, count audits, venue/year confidence analyses, and the machine-readable 77-system master matrix. Copyrighted PDFs, raw library exports, private credentials, repository history, and private working materials are excluded from the package.

Generative-AI tools were used to support title and abstract screening and to assist with language and structural editing of the manuscript. All screening decisions and all AI-assisted manuscript content were reviewed and verified by the authors. The authors retained full responsibility for the extracted evidence, numerical counts, citations, interpretations, and final manuscript. Generative-AI tools were not treated as authoritative sources.

\section{Related Work}
\label{sec:related-work}

The closest external SoKs and surveys are compared earlier in Table~\ref{tab:related-work-comparison}. Broad cryptocurrency and DeFi SoKs organize protocols, applications, attacks, and economic risk \cite{bonneau2015sokBitcoin,werner2022sokDeFi,zhou2023sokDefiAttacks}. Wallet-focused surveys and SoKs organize wallet protocols, authentication factors, attack surfaces, vulnerabilities, incidents, and defenses \cite{karantias2020taxonomyWallets,houy2023walletSlr,homoliak2024sokWallets,erinle2025sokWallets}. These works establish the broader security and financial-technology context, but they do not make recovery semantics--what object or control relation is restored, who authorizes it, and what state remains afterward--the primary unit of comparison.

The retained corpus spans bodies of work that are usually discussed separately: threshold reconstruction and share repair; password-, factor-, hardware-, biometric-, and deterministic-wallet recovery; social and smart-account control transfer; decentralized-identity wallet backup; institutional disaster recovery; and forensic extraction. The shared gap is not the absence of mechanisms, but the absence of a common account of what is restored, who authorizes restoration, and what state remains afterward.

This paper differs from a conventional survey by using recovery semantics as the organizing viewpoint and making those semantics auditable in a system-level matrix. Rather than treating every mechanism as private-key recovery, it distinguishes secret reconstruction from account-control restoration, signing-authority migration, credential re-issuance, asset evacuation, and forensic extraction. That distinction drives the taxonomy, construction model, and agenda.

\section{Conclusion}
\label{sec:conclusion}

Cryptographic recovery in financial technologies is broader than private-key backup. The retained corpus shows mechanisms for reconstructing secrets, regenerating wallet seeds, repairing shares, restoring credentials, delegating recovery, rotating account ownership, evacuating assets, and extracting wallet artifacts. A wallet that reconstructs a seed, a custody protocol that repairs a signing share, a smart account that rotates ownership, and a forensic tool that extracts artifacts are not making the same promise. Recovery-aware financial systems need explicit recovery semantics before they can compare safety, liveness, privacy, usability, and trust, or remain usable after loss without creating unacceptable paths for theft, coercion, surveillance, or exclusion.

\bibliography{main}

\appendix
\section{Full 77-System Master Matrix}
\label{app:master-matrix}

The following tables reproduce the complete master matrix used for the SoK synthesis. The matrix is split into four parts for page layout; the parts are keyed by the Study citation and kept in the same row order.

\begin{landscape}

\subsection{Classification and Mechanism Fields}
\begingroup
\scriptsize
\setlength{\tabcolsep}{2pt}
\renewcommand{\arraystretch}{1.08}

\begin{longtable}
{@{}L{0.05\linewidth}L{0.21\linewidth}L{0.17\linewidth}L{0.17\linewidth}L{0.22\linewidth}L{0.18\linewidth}@{}}
  \caption{Master matrix, part A: classification and mechanism fields.}\label{tab:master-matrix-classification}\\
  \hline
  Study & Financial context & Recovered object & Secret/control axis & Mechanism family & Enrollment/storage \\
  \hline
  \endfirsthead
  \caption[]{Master matrix, part A: classification and mechanism fields. (continued)}\\
  \hline
  Study & Financial context & Recovered object & Secret/control axis & Mechanism family & Enrollment/storage \\
  \hline
  \endhead
  \hline
  \endfoot
  \cite{2} & self custody wallet & original private key & secret restoring & password protected secret sharing & password or factor material; server held protected material \\
  \cite{6} & self custody wallet; non financial transferable pattern & original private key & secret restoring & threshold secret reconstruction & distributed shares or committees \\
  \cite{8} & self custody wallet; DeFi or smart contract application & account control; asset ownership or transfer & control restoring & smart contract control transfer & on chain state or pre authorized transfer \\
  \cite{10} & self custody wallet; smart contract account; institutional custody; DeFi or smart contract application; secure hardware for financial infrastructure & signing authority; account control; asset ownership or transfer & control restoring & smart contract control transfer & on chain state or pre authorized transfer \\
  \cite{11} & institutional custody & original private key & secret restoring & threshold secret reconstruction & distributed shares or committees \\
  \cite{14} & self custody wallet; payment or permissioned blockchain & adversarial key extraction & forensic or adversarial extraction & requirements or vision & analysis framework no single storage \\
  \cite{16} & non financial transferable pattern & seed or root secret & secret restoring & password protected secret sharing & password or factor material; server held protected material \\
  \cite{18} & self custody wallet & signing authority; account control & control restoring & institutional backup or disaster recovery & institutional custody media and procedures \\
  \cite{19} & self custody wallet & seed or root secret; account control & hybrid secret and control restoring & social or guardian recovery & guardians or delegates \\
  \cite{23} & self custody wallet; hardware wallet & signing authority & control restoring & deterministic seed regeneration & user retained seed or memory artifact \\
  \cite{24} & payment or permissioned blockchain & original private key & secret restoring & threshold secret reconstruction & distributed shares or committees \\
  \cite{25} & self custody wallet & original private key & secret restoring & threshold secret reconstruction & distributed shares or committees \\
  \cite{27} & institutional custody; payment or permissioned blockchain; DeFi or smart contract application & key share; signing authority & hybrid secret and control restoring & share repair or refresh & distributed shares or committees \\
  \cite{28} & payment or permissioned blockchain; institutional custody; non financial transferable pattern & key share & secret restoring & share repair or refresh & distributed shares or committees \\
  \cite{29} & self custody wallet, non financial transferable pattern & recovery key & secret restoring & password protected secret sharing & password or factor material; server held protected material \\
  \cite{30} & self custody wallet; hardware wallet & seed or root secret & secret restoring & TEE or HSM assisted recovery & secure hardware or wrapped secret \\
  \cite{31} & self custody wallet; decentralized identity for finance & original private key & secret restoring & threshold secret reconstruction & distributed shares or committees \\
  \cite{34} & self custody wallet; exchange or hosted wallet; payment or permissioned blockchain & original private key & secret restoring & biometric or fuzzy regeneration & biometric source and helper data \\
  \cite{36} & decentralized identity for finance; self custody wallet; non financial transferable pattern & original private key & secret restoring & threshold secret reconstruction & distributed shares or committees \\
  \cite{37} & self custody wallet & asset ownership or transfer & control restoring & social or guardian recovery & guardians or delegates \\
  \cite{38} & payment or permissioned blockchain & original private key & secret restoring & threshold secret reconstruction & distributed shares or committees \\
  \cite{40} & self custody wallet & signing authority; account control; asset ownership or transfer & control restoring & ownership proof or fallback signing & on chain state or pre authorized transfer \\
  \cite{41} & payment or permissioned blockchain; self custody wallet & original private key & secret restoring & password protected secret sharing & password or factor material; server held protected material \\
  \cite{42} & self custody wallet; institutional custody; DeFi or smart contract application; payment or permissioned blockchain & seed or root secret; signing authority & hybrid secret and control restoring & deterministic seed regeneration & user retained seed or memory artifact \\
  \cite{43} & self custody wallet; decentralized identity for finance & original private key & secret restoring & threshold secret reconstruction & distributed shares or committees \\
  \cite{44} & self custody wallet; forensics or asset investigation & original private key; adversarial key extraction & forensic or adversarial extraction & cryptanalytic key extraction & device backup or forensic evidence \\
  \cite{45} & non financial transferable pattern & original private key & secret restoring & threshold secret reconstruction & distributed shares or committees \\
  \cite{46} & self custody wallet & seed or root secret & secret restoring & threshold secret reconstruction & distributed shares or committees \\
  \cite{48} & institutional custody; secure hardware for financial infrastructure & original private key; plaintext or application data & framework or requirements & requirements or vision & analysis framework no single storage \\
  \cite{49} & self custody wallet; payment or permissioned blockchain & original private key & secret restoring & threshold secret reconstruction & distributed shares or committees \\
  \cite{53} & smart contract account; non financial transferable pattern & recovery key; account control; asset ownership or transfer & hybrid secret and control restoring & smart contract control transfer & on chain state or pre authorized transfer \\
  \cite{54} & smart contract account; self custody wallet; DeFi or smart contract application & account control; asset ownership or transfer & control restoring & smart contract control transfer & on chain state or pre authorized transfer \\
  \cite{55} & self custody wallet; hardware wallet; secure hardware for financial infrastructure & original private key & secret restoring & threshold secret reconstruction; TEE or HSM assisted recovery & distributed shares or committees; secure hardware or wrapped secret \\
  \cite{56} & self custody wallet & seed or root secret; signing authority & hybrid secret and control restoring & deterministic seed regeneration & user retained seed or memory artifact \\
  \cite{57} & self custody wallet & original private key & secret restoring & threshold secret reconstruction & distributed shares or committees \\
  \cite{59} & self custody wallet; smart contract account; institutional custody; payment or permissioned blockchain & asset ownership or transfer; account control; signing authority & control restoring & smart contract control transfer & on chain state or pre authorized transfer \\
  \cite{62} & institutional custody & key share & secret restoring & threshold secret reconstruction & distributed shares or committees \\
  \cite{63} & self custody wallet; institutional custody; payment or permissioned blockchain & signing authority; account control & control restoring & ownership proof or fallback signing & on chain state or pre authorized transfer \\
  \cite{66} & self custody wallet & original private key & secret restoring & memory based regeneration & user retained seed or memory artifact \\
  \cite{68} & self custody wallet & seed or root secret; signing authority; account control & hybrid secret and control restoring & deterministic seed regeneration & user retained seed or memory artifact \\
  \cite{71} & self custody wallet; hardware wallet; forensics or asset investigation & original private key; seed or root secret; credential; forensic artifact; adversarial key extraction & forensic or adversarial extraction & forensic artifact recovery & device backup or forensic evidence \\
  \cite{72} & self custody wallet & original private key & secret restoring & biometric or fuzzy regeneration & biometric source and helper data \\
  \cite{73} & self custody wallet & original private key; seed or root secret & secret restoring & social or guardian recovery & guardians or delegates \\
  \cite{74} & self custody wallet & original private key & secret restoring & password protected secret sharing & password or factor material; server held protected material \\
  \cite{75} & self custody wallet; smart contract account & asset ownership or transfer & control restoring & pre signed asset evacuation & on chain state or pre authorized transfer \\
  \cite{76} & forensics or asset investigation & forensic artifact & forensic or adversarial extraction & forensic artifact recovery & device backup or forensic evidence \\
  \cite{79} & self custody wallet & original private key; key share & secret restoring & threshold secret reconstruction & distributed shares or committees \\
  \cite{80} & self custody wallet & key share; seed or root secret; account control & hybrid secret and control restoring & threshold secret reconstruction & distributed shares or committees \\
  \cite{81} & self custody wallet; secure hardware for financial infrastructure; non financial transferable pattern & recovery key & secret restoring & biometric or fuzzy regeneration & biometric source and helper data \\
  \cite{82} & self custody wallet; smart contract account; DeFi or smart contract application & recovery key; account control; asset ownership or transfer & hybrid secret and control restoring & social or guardian recovery & guardians or delegates \\
  \cite{83} & non financial transferable pattern & original private key; key share & secret restoring & threshold secret reconstruction & distributed shares or committees \\
  \cite{84} & self custody wallet & original private key & secret restoring & threshold secret reconstruction & distributed shares or committees \\
  \cite{85} & decentralized identity for finance; institutional custody; non financial transferable pattern & seed or root secret; signing authority & hybrid secret and control restoring & TEE or HSM assisted recovery & secure hardware or wrapped secret \\
  \cite{86} & self custody wallet; hardware wallet; smart contract account & account control & control restoring & passkey or backup authenticator recovery; smart contract control transfer & on chain state or pre authorized transfer; backup authenticator or reissued credential \\
  \cite{87} & hardware wallet; self custody wallet; smart contract account; secure hardware for financial infrastructure & account control; asset ownership or transfer; signing authority & control restoring & smart contract control transfer & on chain state or pre authorized transfer \\
  \cite{89} & self custody wallet; DeFi or smart contract application & original private key & secret restoring & threshold secret reconstruction & distributed shares or committees \\
  \cite{90} & payment or permissioned blockchain & original private key & secret restoring & threshold secret reconstruction & distributed shares or committees \\
  \cite{93} & self custody wallet & seed or root secret & secret restoring & deterministic seed regeneration & user retained seed or memory artifact \\
  \cite{94} & decentralized identity for finance; self custody wallet; non financial transferable pattern & key share; signing authority & hybrid secret and control restoring & threshold secret reconstruction & distributed shares or committees \\
  \cite{95} & self custody wallet & seed or root secret & secret restoring & biometric or fuzzy regeneration & biometric source and helper data \\
  \cite{96} & self custody wallet & signing authority; account control; asset ownership or transfer & control restoring & ownership proof or fallback signing & on chain state or pre authorized transfer \\
  \cite{97} & self custody wallet; payment or permissioned blockchain & seed or root secret; account control & hybrid secret and control restoring & deterministic seed regeneration & user retained seed or memory artifact \\
  \cite{98} & self custody wallet; non financial transferable pattern & original private key & secret restoring & threshold secret reconstruction & distributed shares or committees \\
  \cite{99} & secure hardware for financial infrastructure; non financial transferable pattern & original private key & secret restoring & TEE or HSM assisted recovery & secure hardware or wrapped secret \\
  \cite{101} & institutional custody; exchange or hosted wallet; secure hardware for financial infrastructure & original private key; key share; signing authority; asset ownership or transfer & hybrid secret and control restoring & institutional backup or disaster recovery & institutional custody media and procedures \\
  \cite{102} & non financial transferable pattern; decentralized identity for finance & credential; account control & control restoring & passkey or backup authenticator recovery; credential reissuance or rotation & backup authenticator or reissued credential \\
  \cite{103} & forensics or asset investigation & forensic artifact & forensic or adversarial extraction & forensic artifact recovery & device backup or forensic evidence \\
  \cite{106} & decentralized identity for finance; non financial transferable pattern & credential; plaintext or application data & hybrid secret and control restoring & institutional backup or disaster recovery & institutional custody media and procedures \\
  \cite{107} & self custody wallet & seed or root secret & secret restoring & threshold secret reconstruction & distributed shares or committees \\
  \cite{108} & self custody wallet; DeFi or smart contract application & signing authority; account control & control restoring & password protected secret sharing & password or factor material; server held protected material \\
  \cite{109} & self custody wallet; smart contract account; DeFi or smart contract application & original private key; key share; account control; signing authority & framework or requirements & recovery analysis framework & analysis framework no single storage \\
  \cite{111} & self custody wallet & recovery key; signing authority & hybrid secret and control restoring & biometric or fuzzy regeneration & biometric source and helper data \\
  \cite{112} & self custody wallet; hardware wallet; secure hardware for financial infrastructure & original private key & secret restoring & passkey or backup authenticator recovery & backup authenticator or reissued credential \\
  \cite{113} & payment or permissioned blockchain; self custody wallet & original private key & secret restoring & credential reissuance or rotation & backup authenticator or reissued credential \\
  \cite{114} & self custody wallet & asset ownership or transfer & control restoring & pre signed asset evacuation & on chain state or pre authorized transfer \\
  \cite{117} & decentralized identity for finance & signing authority; key share & hybrid secret and control restoring & threshold secret reconstruction & distributed shares or committees \\
  \cite{118} & self custody wallet; DeFi or smart contract application & original private key; signing authority & hybrid secret and control restoring & deterministic seed regeneration & user retained seed or memory artifact \\
\end{longtable}
\endgroup

\subsection{Recovery Semantics}
\begingroup
\scriptsize
\setlength{\tabcolsep}{2pt}
\renewcommand{\arraystretch}{1.08}

\begin{longtable}{@{}L{0.05\linewidth}L{0.27\linewidth}L{0.20\linewidth}L{0.28\linewidth}L{0.20\linewidth}@{}}
  \caption{Master matrix, part B: recovery semantics.}\label{tab:master-matrix-semantics}\\
  \hline
  Study & Recovery trigger & Failure event class & Authorization method & Post-recovery state \\
  \hline
  \endfirsthead
  \caption[]{Master matrix, part B: recovery semantics. (continued)}\\
  \hline
  Study & Recovery trigger & Failure event class & Authorization method & Post-recovery state \\
  \hline
  \endhead
  \hline
  \endfoot
  \cite{2} & loss of local secret availability with password retained; user initiates reconstruction by contacting t+1 servers & key material or device loss & credential & same or equivalent secret available \\
  \cite{6} & lost device or lost private key / later need to retrieve encrypted backup & device or credential loss migration; key material or device loss & any k confidants return valid shares & same or equivalent secret available \\
  \cite{8} & lost device or lost private key/account credentials & device or credential loss migration; key material or device loss & prior transaction partners vote with three-fifths supermajority; original owner can veto & new control relation or credential \\
  \cite{10} & unavailable signer, custodian refusal/unavailability, or smart-contract paralysis proven by timeout or failed test & participant or service unavailability; institutional disaster or governance change & migration rule satisfied by on-chain challenge-response, Merkle proof of failed test, or TEE-signed transition & new control relation or credential \\
  \cite{11} & enterprise joint authorization need for transaction signing under distributed control & not stated or general loss & per-sub-account CP-ABE access tree plus global t-of-n threshold & same or equivalent secret available \\
  \cite{14} & key leakage, exposed derived key plus higher-level public key, or repeated randomness in stealth-address transactions & not stated or general loss & recipient spend secret authorizes normal spending; no separate authorized recovery workflow stated & analysis only no single post state \\
  \cite{16} & need to retrieve previously shared secret; succeeds only with correct password and at least t+1 honest servers & key material or device loss & password only & same or equivalent secret available \\
  \cite{18} & one or both service units fail; disaster destroys provider-side data & institutional disaster or governance change & Bitcoin 3-of-5 multisignature; user provides more keys in degraded modes & operational access restored or rekeyed \\
  \cite{19} & lost device or device replacement/update of the management device & device or credential loss migration; key material or device loss & user personal information plus signature-based server authentication, with biometric authentication between user and management device & new control relation or credential \\
  \cite{23} & hot-wallet compromise and later need for cold wallet to derive per-transaction signing key from on-chain ID & compromise or theft; key material or device loss & cryptographic authorization by deriving sk ID on the cold wallet and producing a valid signature & same or equivalent secret available \\
  \cite{24} & lost device or terminal damage causing loss of private key & device or credential loss migration; key material or device loss & zero-knowledge-proof-style password knowledge proof plus threshold share return & same or equivalent secret available \\
  \cite{25} & local key data missing on reconnect; storage-node offline or node destruction affecting availability & participant or service unavailability & biometric-derived identity material plus routing ID, authentication info, and nonce-based challenge-response & same or equivalent secret available \\
  \cite{27} & unavailable signer & participant or service unavailability; key material or device loss & threshold verification and interpolation with committee participation and on-chain commitment checks & share or committee refreshed \\
  \cite{28} & unavailable or missing replica share during PBFT execution, view change, or state transfer & participant or service unavailability; key material or device loss & threshold recovery contributions from f+1 replicas with verification & share or committee refreshed \\
  \cite{29} & device loss, failure, or theft & compromise or theft; key material or device loss & password for remote PPSS plus out-of-band authentication for local/social shares & same or equivalent secret available \\
  \cite{30} & lost device or broken wallet motivates proactive backup creation & device or credential loss migration; key material or device loss & user visual verification of vcode on trusted wallet screens plus physical-button confirmation & same or equivalent secret available \\
  \cite{31} & lost key or need to restore wallet secret key & device or credential loss migration; key material or device loss & wallet authenticates user; threshold share release from at least t providers & same or equivalent secret available \\
  \cite{34} & every signing event; motivated by lost or leaked stored keys & key material or device loss & biometric sample sufficiently close to enrollment & same or equivalent secret available \\
  \cite{36} & lost device or key access, forgotten passphrase, hardware corruption, long-term loss of access & device or credential loss migration; key material or device loss & possession of secured private key sK plus threshold subset of correct answers to personal security questions & same or equivalent secret available \\
  \cite{37} & lost device or lost private key / inaccessible original wallet key & device or credential loss migration; key material or device loss & threshold assistants + zero-knowledge user proof + off-chain identity confirmation & new control relation or credential \\
  \cite{38} & lost device or lost private key due to crash, data damage, or force majeure & device or credential loss migration; key material or device loss & CA re-authentication plus identity confirmation by committer peers and user biometric key & same or equivalent secret available \\
  \cite{40} & compromised key & compromise or theft & challenge-response proof using hidden backup key & new control relation or credential \\
  \cite{41} & lost device or lost private key is implied; exact trigger conditions beyond key unavailability are not stated & participant or service unavailability; device or credential loss migration; key material or device loss & password-based authorization using p or p' with threshold server responses & same or equivalent secret available \\
  \cite{42} & lost device or wallet, hardware failure, natural disaster, catastrophic loss; also hierarchy change or key replacement for rekeying & device or credential loss migration; key material or device loss; institutional disaster or governance change & possession of initial seed and master secret rooted certificate authority & same or equivalent secret available \\
  \cite{43} & lost device or key unavailability not formally specified; recovery begins when local secret is unavailable & participant or service unavailability; device or credential loss migration; key material or device loss & biometric identity verification via reconstructed IBV and CBV comparison, then threshold provider response & same or equivalent secret available \\
  \cite{44} & nonce reuse or complementary nonce causing duplicated r & forensic or adversarial recovery; key material or device loss & None; any observer with public blockchain data can perform extraction & artifact or key extracted for investigation \\
  \cite{45} & Not stated; reconstruction is described when access to the encrypted key-protected data is needed, but no explicit loss, compromise, or disaster trigger is defined. & compromise or theft; key material or device loss; institutional disaster or governance change & threshold possession of shares & same or equivalent secret available \\
  \cite{46} & device loss, theft, corruption, or destruction & compromise or theft; key material or device loss & user supplies both ciphertexts and corresponding decryption components, or access to two configured endpoints/accounts & same or equivalent secret available \\
  \cite{48} & missing periodic heart-beat after timeout due to sysadmin unavailability & death or incapacity; participant or service unavailability & Not stated consistently; examples mention any other supplier actor, CEO, regulator, or judge-assisted process; this matters because unclear authorization prevents comparison of abuse resistance and liveness & analysis only no single post state \\
  \cite{49} & lost wallet private key; insufficient primary-threshold peer responses due to failure or malicious behavior & participant or service unavailability; key material or device loss & CA certificate plus biometric key SK & same or equivalent secret available \\
  \cite{53} & lost device or lost main blockchain private key for the original account; only contract-deposited assets are recoverable & device or credential loss migration; key material or device loss & knowledge of answers to security questions plus signature under reconstructed proof key & new control relation or credential \\
  \cite{54} & key loss, key theft, or policy-flagged exceptional transaction & compromise or theft; key material or device loss & digital signature plus DZT security-question authentication or U2F token; optionally 2FA-only in liberal mode & new control relation or credential \\
  \cite{55} & lost/broken/stolen device; lost access to key; lost password & compromise or theft; key material or device loss & possession of threshold number of shares & same or equivalent secret available \\
  \cite{56} & lost device; forgotten password or mnemonic & device or credential loss migration; key material or device loss & possession of stego image plus synchronized address book and user re-entry/reselection of recovery parameters & same or equivalent secret available \\
  \cite{57} & lost device or lost key not stated precisely; evidence supports lost key & forensic or adversarial recovery; device or credential loss migration; key material or device loss & qualified participant subset under general access structure / threshold & same or equivalent secret available \\
  \cite{59} & accidental key loss or transfer to mistyped/unknown address & key material or device loss & claimant initiates commit-reveal-claim; current owner vetoes by proof of possession within challenge window & new control relation or credential \\
  \cite{62} & lost device or lost share; paper states loss of a secret share, including simple loss or extortion-triggered loss of availability & device or credential loss migration; key material or device loss & threshold participation by t helper parties & same or equivalent secret available \\
  \cite{63} & lost key, unreachable signer, or signer unavailability & participant or service unavailability; key material or device loss & remaining online signer engages offline recovery party using pre-distributed encrypted recovery material & new control relation or credential \\
  \cite{66} & lost device or lost private key / inability to use mnemonic backup; exact trigger is loss of private key & device or credential loss migration; key material or device loss & user recalls picture locations and repository returns stored pictures/public key based on user's basic information; repository authentication not stated & same or equivalent secret available \\
  \cite{68} & hardware breakdown or wallet unavailability requiring reconstruction from backed-up master secret & participant or service unavailability; key material or device loss & possession of master secret key (a,b) by the wallet owner & same or equivalent secret available \\
  \cite{71} & law-enforcement search and urgent seizure need; live host access before volatile artefacts disappear & forensic or adversarial recovery & legal search/seizure authority under POCA/PACE, not user authorization & artifact or key extracted for investigation \\
  \cite{72} & lost private key or need wallet access to initiate transaction & key material or device loss & biometric measurement of the user; exact verification process not stated & same or equivalent secret available \\
  \cite{73} & lost device secret in the form of lost private key or mnemonic phrase & device or credential loss migration; key material or device loss & threshold shareholder approval plus email OTP plus fingerprint-based fuzzy vault decoding & new control relation or credential \\
  \cite{74} & lost or forgotten authentication factor & key material or device loss & threshold t-of-n authentication factors via threshold MFKDF & same or equivalent secret available \\
  \cite{75} & lost device or inaccessible original private key; example includes forgotten password for hot wallet access & device or credential loss migration; key material or device loss & pre-authorization via transaction signed with original key at setup; no strong authentication at activation in base design & assets moved to recovery path \\
  \cite{76} & confiscated device investigation where investigators lack wallet address/private key & forensic or adversarial recovery; key material or device loss & legal warrant and, in some cases, exchange or token-issuer cooperation & artifact or key extracted for investigation \\
  \cite{79} & user requests key restoration; periodic scrubbing detects failed or invalid server share & key material or device loss & server-to-server proof-based validation for share checking; user recovery request authentication Not stated & same or equivalent secret available \\
  \cite{80} & lost device share due to storage loss/app uninstall/device change; password change; SIM loss discussed operationally & device or credential loss migration; key material or device loss & user password plus SIM-held data plus server-held encrypted share for lost-share recovery & same or equivalent secret available \\
  \cite{81} & lost device-stored key or lost cryptographic key & device or credential loss migration; key material or device loss & fresh biometric capture plus helper data Q/CP and external salt or identifier & same or equivalent secret available \\
  \cite{82} & lost device or lost private key; compromised key & compromise or theft; device or credential loss migration; key material or device loss & threshold guardian assistance plus biometric verification plus smart-contract check of registered recovery key & new control relation or credential \\
  \cite{83} & epoch transition under node churn and compromise accumulation & compromise or theft; key material or device loss & hierarchical threshold access structure K & same or equivalent secret available \\
  \cite{84} & user needs to use private key; motivated by loss/non-storage of local wallet secret & key material or device loss & WebAuthn/FIDO2 hardware-token authentication plus threshold server verification & same or equivalent secret available \\
  \cite{85} & key loss / system failure & key material or device loss & Not stated; only general MFA, separation of duties, and four-eyes controls are mentioned, which matters because recovery authorization is central to abuse resistance & same or equivalent secret available \\
  \cite{86} & lost device; stolen device; malfunction; compromise of hardware wallet & compromise or theft; device or credential loss migration; key material or device loss & pre-signed ARC by current owner plus TK/passkey signature over new owner address & new control relation or credential \\
  \cite{87} & hardware wallet loss, theft, or malfunction without backup & compromise or theft; key material or device loss & KRC signature verification plus secret-and-commitment consistency checks enforced by registry smart contract & new control relation or credential \\
  \cite{89} & lost device, data damage, or hacking causing loss/leakage of wallet data; user initiates recovery rounds & device or credential loss migration; key material or device loss & threshold participant submissions plus smart-contract verification and user-side checking & same or equivalent secret available \\
  \cite{90} & lost or unavailable private key; reconstruction also threatened by silent or colluding share-holding nodes & participant or service unavailability; key material or device loss & user reconstruction request in a CA-authenticated consortium blockchain setting; exact recovery authorization procedure not stated & same or equivalent secret available \\
  \cite{93} & lost device or device crash of wallet host & device or credential loss migration; key material or device loss & possession of seed or master secret; no separate recovery authorization stated & same or equivalent secret available \\
  \cite{94} & lost device or lost/damaged user secret share in DID wallet & device or credential loss migration; key material or device loss & user initiates by sending new identifier to TTP; secure authenticated channels assumed & same or equivalent secret available \\
  \cite{95} & lost mnemonic phrase & key material or device loss & biometric reproduction via fuzzy extractor; optional additional 2FA for helper-data download is suggested & same or equivalent secret available \\
  \cite{96} & compromised or exposed classical wallet private key; anticipated quantum break of classical signatures & compromise or theft; key material or device loss & user proves knowledge of backup key linked to wallet key via non-interactive zero-knowledge proof, using backup key and stored x & new control relation or credential \\
  \cite{97} & lost credentials, forgotten password, hardware failure, failed authentication, corrupted or compromised wallet & compromise or theft; device or credential loss migration; key material or device loss & Not stated; paper only says authorized Global Ledger users can access the Recovery Ledger & same or equivalent secret available \\
  \cite{98} & lost device or lost private key; unavailable signer or offline fragment holder & participant or service unavailability; device or credential loss migration; key material or device loss & threshold number of verified participating nodes & same or equivalent secret available \\
  \cite{99} & lost device or device switching; later need to retrieve stored key with password & device or credential loss migration; key material or device loss & password-based authentication, with HSM-backed checks and in OPRF-PPKR recovered signing key proof & same or equivalent secret available \\
  \cite{101} & media failure, KMF disaster, media loss/theft, personnel change, threshold change, suspected compromise & compromise or theft; key material or device loss; institutional disaster or governance change & M-of-N treasurer approval, with all N required for initial KeyGen & operational access restored or rekeyed \\
  \cite{102} & lost device or lost primary authenticator & device or credential loss migration; key material or device loss & backup authenticator proves possession of backup key by producing a recovery response/signature over a server challenge tied to a stored recovery credential & new control relation or credential \\
  \cite{103} & discovery or suspicion of cryptocurrency-related artifacts in seized evidence & forensic or adversarial recovery & procedural controls including user validation for private key material, witness requirement, and supervisor notifications & artifact or key extracted for investigation \\
  \cite{106} & lost device and restore onto a new phone from prior encrypted backup & device or credential loss migration; key material or device loss; institutional disaster or governance change & Pod server username plus hashed password, plus separate backup password for local decryption & operational access restored or rekeyed \\
  \cite{107} & lost, stolen, damaged, or destroyed device or hardware wallet; private key loss or hardware failure & compromise or theft; key material or device loss & Not stated clearly; appears to rely on possession of decryption keys, encrypted parts, and correct ordering information & same or equivalent secret available \\
  \cite{108} & forgotten password; lost TOTP device; lost key file; portability need from another device & device or credential loss migration; key material or device loss & email OTP/OOBA plus TOTP and password reset workflow & same or equivalent secret available \\
  \cite{109} & key loss; user unavailability/non-response; agent corruption/collusion affecting composed recovery paths & participant or service unavailability; key material or device loss & threshold shares/components for key reconstruction; guardian approval plus challenge window for key rotation; KELP commit-reveal-claim-challenge with collateral and delays & analysis only no single post state \\
  \cite{111} & biometric unavailability; loss or user death & death or incapacity; participant or service unavailability; key material or device loss & user biometric participation plus wallet-provider helper-key activation; inheritance also uses heir biometric and shared temporal key & same or equivalent secret available \\
  \cite{112} & loss/failure of one MFA factor, especially token; loss of terminal device & key material or device loss & remaining MFA factors plus stored values, with keystore-assisted release of CT-T and suggested out-of-band code & new control relation or credential \\
  \cite{113} & lost private key; inheritance is motivated but workflow details are not stated & death or incapacity; device or credential loss migration; key material or device loss & identity verification by TA plus committee enrollment approval & new control relation or credential \\
  \cite{114} & private key loss; owner disappearance/death & death or incapacity; key material or device loss & identity verification by trusted recovery service provider & assets moved to recovery path \\
  \cite{117} & lost or compromised wallet key share; participant unavailability; key revocation lifecycle events & compromise or theft; participant or service unavailability; key material or device loss & 2-of-3 threshold signing plus registered device checks and TTP key-status verification & same or equivalent secret available \\
  \cite{118} & lost device; smartphone loss, damage, or change; IPFS unavailability & participant or service unavailability; device or credential loss migration; key material or device loss & 24-word mnemonic plus password; smart-contract device authorization; prototype also uses QR scans & same or equivalent secret available \\
\end{longtable}
\endgroup

\subsection{Trust, Threats, and Evidence}
\begingroup
\scriptsize
\setlength{\tabcolsep}{2pt}
\renewcommand{\arraystretch}{1.08}

\begin{longtable}{@{}L{0.05\linewidth}L{0.27\linewidth}L{0.27\linewidth}L{0.16\linewidth}L{0.12\linewidth}L{0.13\linewidth}@{}}
  \caption{Master matrix, part C: trust, threats, and evidence.}\label{tab:master-matrix-evidence}\\
  \hline
  Study & Trust model & Adversary model & Security properties & Validation type & Deployment status \\
  \hline
  \endfirsthead
  \caption[]{Master matrix, part C: trust, threats, and evidence. (continued)}\\
  \hline
  Study & Trust model & Adversary model & Security properties & Validation type & Deployment status \\
  \hline
  \endhead
  \hline
  \endfoot
  \cite{2} & threshold trust in servers; up to t compromised servers reveal nothing, trusted/authenticated initialization required & unauthorized recovery via online password guessing, compromise of up to t servers, and man-in-the-middle control of communication channels & safety, liveness, privacy, non custodiality, robustness & formal proof, security game & conceptual design with formal analysis; no prototype or operational deployment stated \\
  \cite{6} & distributed social trust; secure if fewer than k confidants collude; recovery tolerates up to n-k unavailable helpers & unauthorized recovery via colluding confidants, physical theft, social engineering, and remote compromise avoided by offline storage; formal model not stated & safety; availability; non custodiality; limited accountability; usability tradeoff & prototype; user study; survey or interview & prototype study workflow, not operational deployment \\
  \cite{8} & decentralized social verification by prior transaction partners plus correct smart-contract enforcement; no custodian & fraudulent recovery requester, social engineering, vote buying, Sybil/duplicate-account voter manipulation, 51\% attack & safety; privacy; liveness; availability; non custodiality; accountability; revocability; usability & prototype; simulation & prototype implementation on Hyperledger Composer; no operational deployment stated \\
  \cite{10} & blockchain as censorship-resistant stateful channel plus smart contract or SGX trust anchor & Byzantine corrupted parties, strong network adversary, malicious host/OS against TEE, stale-proof and false-accusation attempts; formal proof for static adversaries & safety; liveness; availability; robustness; accountability; revocability; partial non custodiality & formal proof; prototype; benchmark; gas cost measurement; case study & prototype with Bitcoin testnet deployment evidence; no operational production deployment stated \\
  \cite{11} & trusted owner/master-account for setup and final reconstruction; authorized departmental users for share decryption; standard trust in Bitcoin miners for transaction validation & unauthorized managers, colluding participants, eavesdroppers, transaction tamperers, and possible trading-platform theft as informally discussed & safety, robustness, privacy, availability & prototype, benchmark & prototype only \\
  \cite{14} & recipient protects spend key(s); sender uses fresh randomness; auditor/read-only wallet can monitor but not spend & attacker recovers master private key from derived key plus higher-level public key, or spend key from audit material plus repeated randomness & safety; privacy; availability; robustness; non custodiality & formal proof; security game & analysis of deployed systems and variants; no prototype or operational deployment by authors stated \\
  \cite{16} & threshold trust in n servers with at most t corrupt; secure/authenticated channels and correct setup assumed & up to t corrupted servers, online password-guessing interactions, and network/man-in-the-middle adversary; static corruption in shown theorem & safety, liveness, privacy, availability, non custodiality & formal proof, security game & no evaluation \\
  \cite{18} & user retains 3 keys; two provider service units each store one encrypted key; disaster recovery center restores provider-side state & theft or cracking of provider-stored keys; broader adversary model not stated & safety; availability; robustness & simulation & conceptual design with analytical and simulated evaluation; operational deployment not stated \\
  \cite{19} & honest-but-curious central server plus semi-trusted social-network helpers; secure runtime assumed & server database reader, attacker with access to management device storage, malware/physical theft discussed informally; collusion and coercion not fully stated & safety; privacy; availability; liveness; non custodiality; usability & formal proof; prototype; benchmark & prototype only \\
  \cite{23} & self-custodial hot/cold split; hot wallet may be compromised; cold wallet must remain secret and synchronized; miners only verify signatures & hot-wallet compromise, observation of blockchain keys/signatures, state exposure, related-key attacks & safety; privacy; non custodiality & formal proof; security game; benchmark & conceptual design with formal analysis and Bitcoin transaction cost discussion; no operational deployment stated \\
  \cite{24} & user must keep password secret; enough share-holder nodes must be available and not all malicious & collusion and malicious share holders, network attackers, eavesdropping, replay, DDoS, Sybil; password compromise handling not clearly modeled & safety, liveness, availability, robustness, limited privacy & simulation, prototype, benchmark & prototype only \\
  \cite{25} & decentralized storage and neighbor nodes with threshold availability; no central custodian; exact collusion assumptions not stated & malware theft motivation, fake attacks, replay attacks, source-address inference, and partial node destruction; broader collusion model not stated & safety, liveness, availability, robustness, privacy, non custodiality & simulation, benchmark & prototype deployment not stated; operational deployment not stated; evaluated by experiments and simulations only \\
  \cite{27} & threshold-honest committee under active PPT adversary bounds, synchronous blockchain/network assumptions, and KZG setup trust or at-least-one-honest setup assumption & active adversary corrupting up to threshold-bounded committee members, plus possible KZG setup/assumption failure & safety; liveness; availability; robustness; accountability; non custodiality & formal proof; prototype; benchmark; case study & prototype deployment and lab/mainnet experiments only; no operational deployment stated \\
  \cite{28} & BFT threshold trust with n=3f+1, at most f Byzantine replicas, authenticated channels, PKI, eventual synchrony & Byzantine dealer/client and up to f Byzantine replicas; secrecy against fewer than threshold shares & safety; liveness; privacy; availability; robustness; accountability & formal proof; prototype; benchmark & prototype only \\
  \cite{29} & split trust across local holders, social helpers, remote organizations, cloud storage availability, and recommender honesty & unauthorized recovery via collusion and password guessing; denial of recovery via cloud withholding/deletion or request-limit exhaustion & safety, availability, liveness, privacy, non custodiality, revocability, usability & prototype, benchmark & prototype only \\
  \cite{30} & trusted wallet screens/buttons and hardware storage; untrusted smartphone relay; user-in-the-loop verification & compromised terminal, eavesdropping/sniffing, MITM public-key substitution, wallet theft; compromised wallet internals not stated & safety; non custodiality; availability; usability & prototype; benchmark & prototype only \\
  \cite{31} & trusted wallet endpoint; semi-trustworthy escrow providers; trusted DLT lookup and secure channels; fewer than t providers cannot reconstruct & collusion of fewer than t escrow providers is resisted; broader unauthorized recovery and endpoint compromise model not stated & safety; availability; liveness; limited non custodiality; claimed privacy & prototype; no evaluation & conceptual design with high-level proof-of-concept implementation claim, no operational deployment shown \\
  \cite{34} & trust biometric sensor/authentication device for correct regeneration; do not trust server admin with secret-key custody & malware stealing stored keys; server hacking; malicious administrator; credential leakage in hosted wallet model; template leakage discussed informally & safety; non custodiality; privacy claimed for template leakage; usability and automation tradeoff & prototype; benchmark & prototype on Hyperledger Fabric; no operational deployment stated \\
  \cite{36} & self-service local recovery with no external recovery server; trust in user retention of sK and secrecy/memorability of answers & qualitative discussion of key theft, insecure cloud storage, guessable or publicly inferable answers; no formal adversary model stated & safety; liveness; availability; non custodiality; usability & prototype & prototype integration demo in Ethereum-based SSI application \\
  \cite{37} & non-custodial assistants and modified blockchain nodes are trusted for correct participation; at least one CRS setup participant must destroy randomness; off-chain channels are trusted for user authentication & unauthorized recovery by colluding assistants, malicious users/nodes/assistants, ledger observers, and verification-bypass attackers; endpoint compromise and network/consensus attacks excluded & safety, privacy, non custodiality, availability, accountability & formal proof, security game, prototype, benchmark, simulation & prototype on modified Ethereum EVM/geth, not operational deployment \\
  \cite{38} & hybrid cryptographic and institutional trust in threshold peers, CA authentication, and user-held biometric key & man-in-the-middle attacker, colluding or bribed committer peers, malicious peers returning altered shares & safety, robustness, availability, non custodiality, revocability & formal proof & conceptual design only; no prototype or operational deployment stated \\
  \cite{40} & user-held backup key with verifier acceptance; blockchain/protocol support needed for full on-chain fallback & leaked ECDSA secret key attacker; hot-wallet compromise; quantum/ECDLP break; forgery attempts against eW-OTS+ & safety; non custodiality; revocability; accountability; limited liveness & formal proof; security game; prototype; benchmark & prototype only \\
  \cite{41} & honest-but-curious recovery servers with tolerance for some malicious or faulty nodes in a threshold setting & honest-but-curious insiders, external Dolev-Yao attackers, spoofing, eavesdropping, and collusion attempts & safety; privacy; availability; robustness & no evaluation & conceptual design with analytical comparison only \\
  \cite{42} & centralized trust in cold storage holder of seed and msk; hierarchical trust where ancestors can derive descendant keys; blockchain scripting environment must verify certificates & formal corruption of arbitrary users with descendant compromise, signing oracle access, and prevention of upward privilege escalation; cold storage compromise is catastrophic & safety; availability; robustness; accountability; limited revocability; usability & formal proof; prototype; benchmark & prototype on Bitcoin Cash test blockchain; no operational deployment stated \\
  \cite{43} & trusted wallet plus semi-trustworthy key escrow providers and blockchain-based provider registry & provider compromise and sub-threshold collusion explicitly discussed; broader wallet, biometric, coercion, and DoS threats not stated & safety; privacy; availability; non custodiality; limited robustness; limited revocability & prototype; no evaluation & prototype only \\
  \cite{44} & no trusted third party; safety depends on correct ECDSA nonce generation, and attacker needs only public blockchain access & public-data attacker exploiting duplicated signatures to compute Bitcoin private keys & safety failure; limited privacy discussion; non custodiality context & prototype; case study & prototype analysis on historical Bitcoin blockchain data; no deployed recovery or mitigation system stated \\
  \cite{45} & threshold secret-sharing assumption with distributed cloud resource providers; fewer than threshold shares cannot reconstruct, but provider collusion and malicious behavior are not explicitly modeled & external and internal attackers stealing or hacking keys; brute-force attack discussed; collusion and malicious providers not explicitly modeled & safety, availability, liveness, robustness & simulation, benchmark & prototype-like Matlab simulation only; no operational deployment stated \\
  \cite{46} & at least one of two physically separated storage locations remains uncompromised and available; user trusts wallet software and storage/account access paths & unauthorized recovery by attacker compromising one storage location; factoring and brute-force on random value discussed; compromise of both locations not stated & safety, availability, robustness, non custodiality, usability & formal proof & conceptual design only \\
  \cite{48} & trusted sysadmin; semi-honest bank and non-sysadmin supplier actors; additional recovery actors not formalized & semi-honest insiders and external adversaries; escrow collusion, stale-puzzle recovery, blockchain fork interference, malicious sysadmin noted but not solved & safety; liveness; availability; limited accountability discussion; unclonability desired for heart-beat puzzle & no evaluation & conceptual design / vision only \\
  \cite{49} & user trusted; consortium peers store shadow shares and must meet threshold availability; CA trusted for registration and identity confirmation & malicious peers sending wrong shares; bribed peers; man-in-the-middle interception; external key theft attempts; collusion claimed resisted & safety; liveness; availability; robustness; limited non custodiality; partial revocability; claimed privacy & formal proof & no evaluation \\
  \cite{53} & no trusted third party; trust in Ethereum smart contract execution, client-side cryptographic processing, and secrecy of user-chosen answers & public blockchain observer can read questions and encrypted shares and perform unlimited offline guessing; attacker may initiate malicious recovery if answers are cracked & safety, liveness, non custodiality, limited privacy, limited accountability, partial revocability & prototype; simulation; benchmark; gas cost measurement & prototype evaluation only; no operational deployment stated \\
  \cite{54} & smart-contract-enforced policy with threshold committee or hardware token; no central trusted authority stated & stolen key/credentials, corruption of up to t-1 committee parties, offline guessing of low-entropy answers, replay, malformed responses, denial of service & safety; liveness; privacy; availability; robustness; accountability; usability & formal proof; security game; prototype; benchmark; gas cost measurement & prototype deployment on Ethereum only; no operational deployment stated \\
  \cite{55} & trust in Intel SGX enclave and implementations; remote server stores encrypted share; malicious parties fewer than threshold & device/software compromise, keylogging, side-channel attacks, memory exposure of printable share, MITM on server communication, SGX attacks & safety; availability; liveness; non custodiality; qualitative privacy & prototype & prototype implementation only \\
  \cite{56} & user-retained recovery artifacts and local recomputation; no trusted third party for signing authority; cloud storage availability/integrity assumptions are informal & dictionary attacker; blockchain linkability observer; physical theft/observation and malware mentioned informally; cloud compromise and metadata tampering not fully modeled & safety; privacy; availability; liveness; non custodiality; usability & prototype; benchmark; simulation & prototype evaluation only \\
  \cite{57} & distributed participants must retain and provide shares; forbidden subsets learn nothing; setup/model trust not fully stated & unauthorized recovery by forbidden subsets modeled; broader collusion, coercion, malware, and stolen-share threats not stated & safety, liveness, non custodiality & formal proof, simulation, benchmark & conceptual design with lab evaluation; no operational deployment stated \\
  \cite{59} & relies on on-chain enforcement plus synchrony and no prolonged censorship during t2; no secret escrow helper & unauthorized claimers, front-running, random testing, heuristic targeting, deanonymization, short-range and long-range censorship, key destruction attack & safety; liveness; availability; non custodiality; accountability; limited privacy; some revocability in key-rotation settings & prototype; case study & prototype-level Diem implementation noted; no operational deployment stated \\
  \cite{62} & cryptographic trust in Shamir sharing, commitments, secure channels, and at most t-1 static malicious corruptions; active availability of t helpers required & malicious static PPT adversary corrupting up to t-1 parties in a partially synchronous network & safety, privacy, availability, non custodiality & formal proof, security game, prototype, benchmark & prototype only; no operational deployment stated \\
  \cite{63} & trustless setup with non-collusion assumption; secure if sub-threshold parties collude and one helping party remains honest; recovery depends on offline party cooperation & adaptive malicious adversary corrupting at most t-1 players; forgery resistance under Schnorr, non-malleable commitment, DDH, and IND-CPA assumptions & safety; liveness; availability; non custodiality; robustness & formal proof; security game; simulation & conceptual design with formal analysis only \\
  \cite{66} & user memory plus repository and map application availability/integrity; repository security/authentication not stated & unauthorized recovery via repository attack, brute-force GPS guessing, and inference of locations from picture content; many practical threats not stated & safety, usability, limited non custodiality, liveness & prototype; user study; benchmark & prototype deployment in a mobile wallet application; operational deployment not stated \\
  \cite{68} & non-custodial single-owner trust; security relies on secrecy of master spend component and secure backup of master secret & chosen-message adversary with corruption of arbitrary derived signing keys and master secret view key; privacy adversary attempts linkage of derived keys/signatures & safety; privacy; non custodiality; limited usability & formal proof; security game; prototype; benchmark & prototype only \\
  \cite{71} & institutional trust in investigators and evidence handling; operational trust in forensic tools, cloud GPUs, and blockchain APIs & offline dictionary attacker using seized host artefacts; target wallet owner may resist through encryption, hardware-wallet separation, or app termination & availability; accountability; limited safety via forensic integrity measures; weak privacy & prototype; benchmark; case study & prototype and lab evaluation only \\
  \cite{72} & self-service recovery relying on user biometric data and trusted local software; no guardians or custodians stated & unauthorized key guessing and general unauthorized access mentioned; formal adversary model not stated & safety, liveness, non custodiality; privacy and robustness claimed informally but not formalized & prototype, benchmark & conceptual design with lab-style experimental evaluation; no operational deployment stated \\
  \cite{73} & threshold social helpers plus smart contracts, with residual wallet admin trust for Public contract deployment and OTP signing & unauthorized recovery with fewer than threshold shares is addressed; broader collusion, coercion, email compromise, and recovery-endpoint compromise are not stated & safety; privacy; availability; robustness; non custodiality; accountability; usability & prototype; benchmark & prototype on Ganache local Ethereum test network, not operational deployment \\
  \cite{74} & no trusted third party or committee; untrusted public storage for public parameters; security depends on possession of threshold factors & unauthorized recovery via compromised factors, brute-force attacks, storage-layer DoS, temporary eclipse attack, factor compromise such as SIM swapping & safety, liveness, availability, non custodiality, usability & prototype & prototype Ethereum web wallet validated on testnet and mainnet, not stated as operational deployment \\
  \cite{75} & user trusts cold storage of backup key, smart contract correctness, and availability/correctness of Chainlink, external adapter, and IPFS & unauthorized activation, backup-key theft, destination rewriting, forged emergency contract, privacy leakage from exposed recovery data & liveness; non custodiality; accountability; partial privacy, with weakened safety due to unauthenticated activation & prototype; benchmark; gas cost measurement & prototype on Goerli testnet, not stated as operational deployment \\
  \cite{76} & investigators need lawful device access and accurate forensic analysis; some cases require exchange or issuer cooperation & criminal wallet users and encrypted/hidden wallet artifacts; false positives in artifact search; broader misuse threats not stated & liveness, availability, accountability & case study & lab-style case studies on Windows 10 wallets; operational deployment not stated \\
  \cite{79} & distributed backup servers plus system manager and recruitable replacement server; threshold availability required & share tampering, unavailability, and plaintext leakage during repair explicitly considered; colluding servers and unauthorized requester model Not stated & safety; liveness; privacy; availability; partial non custodiality goal & prototype; benchmark; simulation & prototype only \\
  \cite{80} & trusted central server, trusted SIM storage, trusted device/Android KeyStore, assumption that at least 2 of 3 entities are available & informal discussion of brute force, MITM, dictionary attack, keylogger, and DoS; collusion and insider compromise not stated & safety; availability; liveness; robustness; partial non custodiality; usability & no evaluation & conceptual design only; no prototype or deployment stated in supplied evidence \\
  \cite{81} & user biometrics must remain reproducible; helper data storage on device/server/TTP/HSM must remain available; secure communication channel assumed & malware, brute force, ciphertext-only, and known-text attacks considered; communication attacks excluded; spoofing/coercion/helper-data tampering not fully modeled & safety; privacy; robustness; revocability; usability & prototype; benchmark & prototype and lab evaluation; blockchain use case conceptual only \\
  \cite{82} & user-selected guardians with k-of-n threshold; trusted smart-contract enforcement; secure communication channels; biometric mechanism assumptions & unauthorized recovery via guardian collusion or eavesdropping is mentioned informally; no formal adversary model stated & safety; liveness; availability; privacy; robustness; non custodiality & prototype; benchmark; gas cost measurement & prototype only \\
  \cite{83} & no TTP; threshold trust in hierarchical committee, trusted user setup server, blockchain commitment availability, reliable channels & active computationally bounded adversary with threshold-bounded node corruption and arbitrary deviation by corrupted nodes & safety, liveness, availability, robustness, privacy, partial non custodiality & formal proof, prototype, benchmark & prototype only \\
  \cite{84} & threshold trust in distributed key custody servers plus trusted hardware authenticator; client compromise assumptions not stated & unauthorized recovery requires compromise of authentication factors and enough servers; replay addressed; broader endpoint/coercion model not stated & safety, liveness, availability, usability & prototype, benchmark, simulation & prototype only \\
  \cite{85} & HSM-rooted institutional trust; external storage must preserve wrapped-seed availability/integrity; privileged operators constrained by organizational controls & physical and side-channel attacks on HSM, cryptanalytic attacks on wrapped seeds, network attacks, insider abuse, DoS, social engineering, storage attacks, quantum threat & safety; liveness; availability; privacy; accountability; revocability; usability & no evaluation & conceptual design only; prototype and performance testing are future work \\
  \cite{86} & trusted hardware wallet secure environment; trusted smart account and registry logic; secure ARC storage; secure and reliable passkey provider; secure GUI; blockchain execution & unauthorized recovery via ARC plus TK compromise; DoS via stolen ARC registration; replay; brute force; MEV; intercepted on-chain data considered & safety; liveness; availability; non custodiality; revocability; usability & prototype; benchmark & prototype deployment on Ethereum Sepolia testnet, not stated as operational deployment \\
  \cite{87} & user must securely store KRC and secret; hardware wallet secure element and RNG trusted; registry and smart account contracts trusted for correct enforcement; blockchain may be adversarial but eventually processes transactions & MEV/front-running/reordering/censorship adversary; DoS via leaked KRC or random entries; brute force against 256-bit secret; replay after reveal; theft of one or both recovery artifacts & safety; liveness; privacy; availability; non custodiality; accountability; robustness & prototype; benchmark & prototype on Ethereum Sepolia testnet, not operational deployment \\
  \cite{89} & mixed cryptographic and incentive-based threshold model; no trusted nodes claimed for escrow/recovery, but setup uses administrator and multi-server generation & rational or malicious participants, false uploads, withholding, and collusion; endpoint compromise and smart-contract bugs not explicitly modeled & safety; liveness; availability; robustness; non custodiality; accountability & formal proof; simulation; benchmark; gas cost measurement & prototype-style smart-contract simulation only; no operational deployment stated \\
  \cite{90} & threshold trust in subsecret-holding nodes plus smart-contract incentive enforcement and consortium-chain institutional trust & silent, selfish, colluding, or false-share-providing nodes; broader unauthorized recovery model not stated & safety, liveness, availability, robustness, accountability, partial non custodiality & simulation, prototype, benchmark & conceptual design with simulation and prototype-like implementation clues; no operational deployment stated \\
  \cite{93} & user and cold wallet must protect root secret; hot wallet may be compromised; cryptographic trust in lattice assumptions and QROM & quantum adversary with ledger observation and hot-wallet compromise; seed-compromise model not stated & safety, privacy & formal proof, prototype, benchmark & prototype-style implementation and lab benchmarking; no operational deployment stated \\
  \cite{94} & TTP and backup device are trusted to be available and provide correct recovery contributions; secure channels assumed & theft of up to t-1 threshold secrets discussed; unauthorized recovery, helper collusion, and helper compromise not explicitly modeled & safety; liveness; availability; robustness; privacy & prototype; benchmark & prototype only \\
  \cite{95} & user plus external storage availability; confidentiality relies on fuzzy extractor assumptions and biometric device/sample integrity & attacker obtains stored backup artifact and attempts unauthorized recovery by guessing or spoofing biometrics; broader compromise model not stated & safety, availability, non custodiality, usability & prototype, benchmark & prototype only \\
  \cite{96} & non-custodial user-held backup key; trust in cryptographic assumptions and verifier acceptance; no guardians or custodians & attacker steals classical secret key, forges signatures, may be quantum-capable, and may replay or clone non-interactive proofs & safety; non custodiality; privacy; limited usability; post-quantum-oriented soundness claims & formal proof; prototype; benchmark & prototype only \\
  \cite{97} & trust in wallet software, Recovery Ledger access control, and PoA ledger/validator infrastructure & unauthorized access and brute-force guessing discussed; broader compromise and insider model not stated & safety; availability; usability & prototype; simulation; benchmark & prototype-level lab evaluation only \\
  \cite{98} & elected controller node plus threshold of participating nodes; blockchain publication for verification; no fully specified user-authentication step & malicious controller, malicious participating nodes, collusion, impersonation, offline nodes & safety; availability; liveness; robustness; accountability; limited privacy claim & prototype; benchmark; simulation & prototype only \\
  \cite{99} & single server with internal HSM; Lev-1 server corruption, Lev-2 file leakage with trusted HSM core, Lev-3 full server/HSM corruption & malicious server, file leakage, full server/HSM corruption, online guessing, offline guessing after leakage, replay/rerouting/key-planting attacks & safety; privacy; availability; liveness; robustness; usability & formal proof; benchmark & conceptual design with formal analysis and efficiency comparison; no operational deployment stated \\
  \cite{101} & trusted operating terminal integrity, trusted physical KMF controls, threshold trust in treasurers, organizational procedures & insider collusion, media theft/swapping, malware, side channels, transaction tampering, MITM during KMF duplication, malicious software exfiltration & safety; liveness; availability; accountability; usability & prototype; case study & prototype mockup on Bitcoin Testnet; no operational deployment stated \\
  \cite{102} & user-held primary and backup authenticators plus server-stored recovery credential; authenticated registration channel and secure CTAP client-token communication assumed & active network/protocol adversary and malicious server considered; quantum adversary motivates pq design; replay and unlinkability attacks modeled & safety via recovery authentication; privacy via unlinkability; limited availability due to single backup token assumption; partial non custodiality & formal proof; security game; benchmark & conceptual design with formal analysis; no full prototype deployment \\
  \cite{103} & institutional and cloud-service trust; investigators, witnesses, supervisors, and SaaS environment must be trusted & internal bad actors and operational misuse explicitly mentioned; broader cloud and key-misuse adversaries not stated & accountability, usability, some operational availability; no formal safety or privacy proof stated & prototype; case study; operational deployment & cloud-based SaaS in use by at least one U.S. law-enforcement team; on-prem containerized version planned \\
  \cite{106} & user-managed credentials plus trusted Pod server/Pod provider availability and access control; no threshold or guardian model & spoofing, tampering, repudiation, information disclosure, denial of service, and elevation of privilege under STRIDE; endpoint compromise and stolen-password scenarios not stated & safety, privacy, availability, accountability, usability, non custodiality & formal proof; prototype & prototype only \\
  \cite{107} & self-managed user trust model with distributed storage locations; no guardians or custodian & attacker targets online wallets or storage, may compromise one server, guess split count, storage locations, ordering, or brute-force seed phrase & safety, availability, non custodiality; limited usability discussion & prototype, benchmark & prototype only \\
  \cite{108} & non-custodial client-side key derivation with dependence on user devices, email channel, TOTP app, and public parameter storage; exact collusion threshold not stated & informal threats include credential stuffing, offline brute force, custodial compromise, malware/browser compromise, seed phrase exposure, and multi-device compromise; no formal adversary model & non custodiality; usability; liveness; availability; partial safety claims & prototype; user study & prototype deployment in user study; no operational deployment stated \\
  \cite{109} & rational profit-motivated adversary; some non-collusion and distinct-provider assumptions; smart-contract enforcement; user responsiveness matters & colluding compromised agents seeking profitable unauthorized recovery, draining, blackmail/refusal, DoS, and front-running & safety; liveness; availability; robustness; non custodiality; accountability; usability & simulation; case study & conceptual analysis framework with PRISM model checking; no operational deployment evidenced \\
  \cite{111} & user plus wallet provider and backup data manager availability; provider stores helper key but should not recover signing key unilaterally & one corrupted signing party in security experiments; wallet provider and backup data manager alone or colluding should not recover or sign without user involvement & safety; non custodiality; privacy; availability; revocability; usability & formal proof; prototype; benchmark & prototype only \\
  \cite{112} & user retains private key secrecy; keystore assists recovery by storing/releasing factor table; trust in token, biometric liveness, and backup device availability & device compromise, keystore/HSM insider attack, interception of recovery material, biometric spoofing, side-channel leakage, physical coercion & safety; non custodiality; privacy; availability; robustness; usability & prototype; benchmark & prototype-like lab evaluation only; operational deployment not stated \\
  \cite{113} & fully trusted TA plus semi-honest committee with limited malicious members detectable & external adversary may corrupt a small number of users or committee members to obtain identity/private key, evade tracing, or forge/impersonate & safety; privacy; accountability; availability; liveness; robustness; reduced escrow style non custodiality claim; usability & formal proof; prototype; benchmark & prototype only, with future real-world validation stated \\
  \cite{114} & trusted third-party RSP with bank/government-grade institutional trust; RSP availability required & unauthorized recovery and transaction hijacking mentioned informally; malicious RSP, insider abuse, and impersonation not stated & safety, liveness, availability, non custodiality, usability & prototype, benchmark & prototype deployment on Bitcoin Testnet, not operational deployment \\
  \cite{117} & wallet owner, TTP share holder/auditor, and backup node must satisfy threshold assumptions; collusion assumptions not clearly stated & replay/reuse, tampering, revoked-key use, unauthorized device participation, phishing/wallet compromise; helper collusion not clearly modeled & safety; availability; robustness; accountability; revocability; limited privacy & prototype; benchmark & prototype implementation and lab-style experiments; no operational deployment stated \\
  \cite{118} & non-custodial user-controlled recovery anchored in smartphone app, mnemonic/password, smart contract, blockchain, and IPFS; strong OS security assumption & computationally bounded attacker; theft of smartphone or mnemonic, interception/tampering attempts, and key-linkage attacks are discussed; endpoint compromise and coercion are largely absent & safety; privacy; availability; non custodiality; limited revocability; integrity & prototype; benchmark; gas cost measurement & prototype deployed on Sepolia with IPFS test setup; no operational deployment stated \\
\end{longtable}
\endgroup

\subsection{Privacy, Usability, and Limitations}
\begingroup
\scriptsize
\setlength{\tabcolsep}{2pt}
\renewcommand{\arraystretch}{1.08}

\begin{longtable}{@{}L{0.05\linewidth}L{0.20\linewidth}L{0.16\linewidth}L{0.14\linewidth}L{0.23\linewidth}L{0.22\linewidth}@{}}
  \caption{Master matrix, part D: privacy, usability, and limitations.}\label{tab:master-matrix-limitations}\\
  \hline
  Study & Privacy discussion & Usability evidence & Post-quantum relevance & Main limitation & SoK relevance \\
  \hline
  \endfirsthead
  \caption[]{Master matrix, part D: privacy, usability, and limitations. (continued)}\\
  \hline
  Study & Privacy discussion & Usability evidence & Post-quantum relevance & Main limitation & SoK relevance \\
  \hline
  \endhead
  \hline
  \endfoot
  \cite{2} & OPRF hides password input, but broader recovery-metadata privacy is not stated & no evaluation & Not stated & reduced default verifiability weakens robustness and server fault identification; no empirical usability or deployment evidence & proof-backed server-assisted wallet secret recovery using password plus threshold servers \\
  \cite{6} & limited; protects key confidentiality but no formal analysis of helper identity or recovery-attempt leakage & user study; survey or interview & Not stated & slow recovery with weak external validity and no revocation or formal abuse model & empirical evidence for offline social secret-sharing recovery of private keys relevant to self-custody backup \\
  \cite{8} & per-voter encrypted proofs, but public sos and unenforceable encryption create leakage risk & no user study; claimed no pre-loss setup, but user must recall transaction details and recruit at least five prior partners & Not stated & vulnerable to social engineering, vote buying, and transaction-partner manipulation, with no formal security proof or user validation & direct example of non-custodial account recovery via asset transfer to a new account rather than private-key reconstruction \\
  \cite{10} & limited; public on-chain challenge-response and possible on-chain encrypted authentication material, no formal privacy treatment stated & no user study; only performance, fee, gas, and testnet demonstrations & Not stated; this matters because the paper's security assumptions are classical and the SoK tracks whether recovery paths depend on non-post-quantum signatures or hardware assumptions. & recovery improves but does not eliminate safety/availability tradeoffs, and practical Bitcoin recovery is largely dependent on TEE assistance while on-chain smart-contract proofs are costly & strong example of recovering control by dynamic authorization migration rather than reconstructing lost keys \\
  \cite{11} & informal attribute-based identity privacy only & no user study & Not stated & reconstructs the full original private key at the owner and does not address loss, compromise recovery, rotation, or revocation & enterprise Bitcoin threshold key reconstruction with attribute-gated share access; relevant as a governance-oriented recovery pattern rather than user loss recovery \\
  \cite{14} & yes; focuses on receiver unlinkability/anonymity, but notes later output merging can break privacy & no evaluation & Not stated & mostly analyzes attacker key recovery and wallet design tradeoffs rather than legitimate user recovery or post-compromise restoration & supporting wallet key-management paper for privacy, delegated monitoring, and compromise-resilient custody design \\
  \cite{16} & password and secret hidden from up to t corrupt servers; metadata leakage during recovery not stated & no evaluation & Not stated & no deployment or usability evidence, and security relies on ROM plus OMDH/T-OMDH assumptions & threshold password-only secret recovery primitive transferable to wallet and key-backup recovery \\
  \cite{18} & Not stated & Not stated & Not stated & no user-side key-loss recovery and no formalized backup integrity or adversary model in supplied evidence & provider-side wallet disaster recovery restoring multisig service state and practical signing capability \\
  \cite{19} & claims blind wallet recovery so server and helpers do not learn master password or exact shares, but metadata leakage is not formally stated & no user study; only claimed usability plus performance measurements and framework comparison & Not stated & recovery depends on a central server and enough social helpers, under strong secure-environment assumptions and limited collusion/coercion analysis & direct cryptocurrency wallet recovery via server-assisted social threshold reconstruction of a master key \\
  \cite{23} & forward unlinkability for pre-breach keys; privacy can fail if chaincode or poorly chosen IDs reveal linkage & no evaluation & Not stated & not a user-facing recovery protocol and requires strong synchronization/secure-erasure assumptions for forward unlinkability & formal wallet-security framework for compromise-resilient deterministic hot/cold wallets, useful as recovery-adjacent custody analysis rather than direct recovery \\
  \cite{24} & limited; focuses on not revealing password during recovery, broader metadata privacy not stated & no user study & mentions quantum computing as a threat, but no post-quantum recovery design is provided & security and successful recovery depend heavily on memorized password secrecy, with no revocation or compromise-rotation workflow & direct blockchain private-key recovery for token accounts using password-protected threshold secret sharing \\
  \cite{25} & limited; anonymous communication hides backup source address, but broader recovery metadata leakage is not stated & no user study & Not stated & no formal security model/proof, unclear collusion assumptions, and no discussion of biometric failure or post-recovery rotation & decentralized non-custodial wallet key recovery via threshold shares and network-based path/share retrieval \\
  \cite{27} & informal discussion of anonymity leakage from direct off-chain communication; no formal privacy model stated & no user study & Not stated & infrastructure-level share continuity rather than explicit end-user recovery, with trusted-setup-dependent main path and weaker fallback threshold & dynamic committee-based recovery of threshold key control for blockchain custody under churn \\
  \cite{28} & threshold hiding against up to f Byzantine servers, but dealer secret may be exposed if dealer is proven dishonest & no evaluation & Not stated; the notes identify KZG and Pedersen style components but do not discuss post-quantum implications, which matters for long-term financial infrastructure planning. & recovery is system-internal replica share repair, not end-user account or wallet recovery, and no revocation or compromise-remediation workflow is stated & strong example of threshold share recovery for BFT financial infrastructure, but indirect for consumer custody recovery \\
  \cite{29} & password hidden from servers; cloud data alone insufficient; broader privacy formalization not stated & no user study & Not stated & cryptographic robustness omitted, with reliance on user reporting and recommender flagging & split-trust recovery-key reconstruction for self-custody wallet access recovery \\
  \cite{30} & seed confidentiality in transit discussed; broader recovery privacy not stated & no user study; only implementation flow and convenience claims & Not stated & no revocation or compromise-response workflow after duplicating the same master seed onto two wallets & direct hardware-wallet seed backup and recovery mechanism for self-custody \\
  \cite{31} & partial, not formalized & no user study & Not stated & no substantive security or usability evaluation and strong trust in the wallet endpoint & concrete wallet secret-key recovery design using threshold escrow, transferable to financial wallet recovery \\
  \cite{34} & public template claimed hard to invert, but no formal privacy evidence stated & benchmark only; biometric required for every signature; automatic transactions unsupported & Not stated & no support for automatic transactions or smart contracts because biometric input is always required & recovery-adjacent custody design that regenerates signing keys from biometrics instead of storing them \\
  \cite{36} & qualitative only; answers not stored or transferred in SSI workflow, but question-answer data may be guessable from public information & no user study & Not stated & security-usability tradeoff relies on low-entropy security-question answers and lacks formal security and usability validation & explicit reconstruction of the same blockchain private key via self-service partial-knowledge recovery \\
  \cite{37} & hides assistant identities and threshold via zk proofs, but metadata inference may require fake assistants/addresses & no user study & Not stated; this matters because the scheme relies on conventional zk and elliptic-curve components whose long-term recovery security may change under quantum threats. & requires protocol/EVM modifications and trusted off-chain assistant authentication, with weak evidence on real-world usability & privacy-preserving wallet recovery by asset transfer rather than key reconstruction \\
  \cite{38} & anonymity claimed but not formally defined; biometric privacy implications not stated & no user study & Not stated & no empirical validation and incomplete treatment of biometric, availability, and post-recovery operational issues & direct private-key recovery mechanism for permissioned financial blockchain accounts using verifiable threshold shares \\
  \cite{40} & proof-of-ownership is not zero-knowledge; broader privacy leakage not stated & no user study & yes, uses hash-based fallback to address possible quantum break of ECDLP & does not directly restore stolen funds and full fallback requires blockchain/protocol changes; base proof is one-time & wallet recovery by ownership proof and fallback signing after key leakage rather than secret reconstruction \\
  \cite{41} & limited to confidentiality of secret shares and password-protected values; no formal privacy definition & no user study & Not stated & no prototype or formal proof, and no revocation or compromise-recovery path is stated & direct example of original private-key reconstruction via password protected secret sharing in blockchain recovery \\
  \cite{42} & public derivation leaks key relationships and prevents unlinkability in that mode; alternative unlinkability mode discussed informally & no user study & Not stated & recovery depends on preserved seed or msk, and compromise of cold storage is catastrophic with no stated fallback for seed loss & direct example of deterministic wallet key regeneration and organizational rekeying tradeoffs in financial wallets \\
  \cite{43} & local biometric verification shares only comparison result; encrypted bundles and threshold sharing reduce single-custodian exposure & no evaluation & Not stated & no formal security proof and incomplete treatment of cheating, biometric threats, auditability, and wallet-compromise assumptions & concrete wallet key recovery design reconstructing the original secret via threshold escrow with privacy-preserving authorization \\
  \cite{44} & public blockchain transparency enables extraction analysis; no formal privacy treatment & Not stated; paper is not a legitimate user recovery design & Not stated & not a legitimate recovery protocol and no mitigation or post-compromise recovery design is evaluated & attacker-centric evidence that public signature artifacts can reveal wallet private keys under nonce misuse \\
  \cite{45} & Not stated beyond high-level privacy/security motivation; no privacy formalization & no evaluation & Not stated & no formal security proof or explicit malicious-provider/collusion model, and no finance-specific recovery scenario & transferable threshold key reconstruction pattern for custody and backup, but only supporting evidence rather than direct fintech evidence \\
  \cite{46} & limited; confidentiality of seed phrase discussed, broader metadata privacy not stated & no evaluation & Not stated & no formal security model or empirical evaluation, and no handling of both-location compromise or one-location unavailability & direct self-custody seed recovery mechanism that reconstructs the original wallet seed phrase from two separated digital stores \\
  \cite{48} & limited to keeping key hidden during normal operation; recovery privacy not formalized & no direct usability evidence & Not stated; this matters because puzzle hardness assumptions and long-term recovery security may change under quantum-capable adversaries & no concrete heart-beat puzzle construction, proof, or implementation; authorization and malicious-sysadmin handling remain unresolved & institutional financial continuity and outsourced operational key recovery requirements \\
  \cite{49} & claims anonymous recovery behavior, but no formal privacy definition or proof in evidence notes & no user study & Not stated & no prototype or empirical usability/deployment evidence; secure transmission and overhead left unresolved & Direct wallet private-key recovery mechanism for consortium blockchain using dual-threshold secret sharing with biometric-protected shares \\
  \cite{53} & yes; public on-chain questions, encrypted shares, and balances create linkage and targeting risks & no user study; paper informally argues security questions are easier to remember than passwords & Not stated; this matters because the paper does not discuss whether its cryptographic components remain suitable under post-quantum assumptions & recovery depends on weak user-chosen security questions and public on-chain data enables unlimited offline guessing; recovery only covers contract-held assets & transferable smart-contract asset recovery pattern that restores contract-level control without recovering the original private key \\
  \cite{54} & DZT aims to reveal only success/failure, but broader on-chain metadata privacy is not stated & no user study & Not stated & recovery restores fund control by transfer to a fresh address rather than reconstructing the original key, and the unsigned recovery tradeoff is not deeply analyzed & policy-based smart-contract recovery via control transfer after key loss, with concrete threshold and hardware-2FA variants \\
  \cite{55} & qualitative only; encrypted remote storage claimed private, but plaintext paper share and plaintext address storage create exposure & no user study & Not stated & strong SGX trust assumptions plus possible MITM and plaintext share exposure, with no formal evaluation shown & direct wallet private-key recovery mechanism for self-custody using TEE plus secret sharing \\
  \cite{56} & per-transaction fresh keys reduce linkability, but cloud-stored stego image and address book may leak metadata; no formal privacy analysis & no user study; only indirect evidence from framework comparison and password/randomness tests & Not stated & recovery depends on stego image plus synchronized address-book state and lacks formal analysis of cloud compromise, tampering, and revocation & Bitcoin self-custody recovery design that reconstructs root secret and derived keys without daily backups \\
  \cite{57} & limited to secrecy against forbidden subsets; broader metadata/privacy not stated & no user study & Not stated & no realistic wallet integration or full adversary/lifecycle model, and exact key recovery from noisy image reconstruction is unclear & direct private-key recovery mechanism for blockchain, but mostly a transferable secret-sharing construction rather than a deployed financial recovery system \\
  \cite{59} & yes, but informal and limited; reveal and side channels can expose target and claimer links & no user study; claimed simplicity from no prior backup, but monitoring and long delays are required & Not stated & safety depends on timely owner challenge and strong anti-censorship assumptions; no recovery for compromise cases & reactive non-custodial recovery by delayed on-chain control transfer rather than secret reconstruction \\
  \cite{62} & protects helper shares and global secret during recovery, but broader metadata privacy is not stated & no evaluation & Not stated; this matters because the protocol relies on discrete-log-based elliptic-curve assumptions & no robustness or fairness, static-only corruption model, and no revocation or rekeying after recovery & direct threshold-share recovery for blockchain key management \\
  \cite{63} & standard-signature indistinguishability stated; no formal privacy model for recovery metadata & no empirical usability evidence & Not stated & recovery can be blocked by malicious or DoS-targeted offline recovery party; no empirical deployment evidence & clear recovery-by-fallback-signing pattern for crypto-asset custody that restores signing capability rather than the original private key \\
  \cite{66} & yes; pictures may reveal private information and locations, and landmarks/faces may aid inference & user study & Not stated & repository security/authentication is out of scope despite being central, and picture-based recovery leaks privacy and depends on autobiographical secrecy & direct example of wallet original private key recovery via memory-based regeneration rather than social, biometric, or threshold methods \\
  \cite{68} & formal unlinkability of derived verification keys and signatures; auditor with view key can see wallet activity but cannot spend & no user study & Not stated & recovery is only backup-based wallet reconstruction, with no social recovery, threshold recovery, or post-compromise rotation/revocation workflow & supporting mechanism for self-custody wallet recovery via root-secret reconstruction combined with key-insulated compromise containment \\
  \cite{71} & broad host-data extraction acknowledged, but no formal privacy model or minimization analysis & no validated user study; only aspirational claim of 2-hour triage by relatively unskilled technician & Not stated & recovery success is highly state-dependent and often requires live artefacts plus weak passwords; Ledger seed not recovered from examined host artefacts & forensic seizure and adversarial wallet-secret extraction evidence, not end-user account recovery \\
  \cite{72} & DNA biometric dependence is central, but storage, leakage, and revocation of biometric data are not stated & no user study; authors note biometric error rate may hinder practical use & Not stated & biometric error rate and underspecified practical deployment and authorization details & direct private-key recovery proposal for blockchain wallets, useful as a biometric-assisted self-custody recovery pattern with major threat-model and validation gaps \\
  \cite{73} & fingerprint template not stored directly; encoded fingerprint with chaff points is encrypted, but broader metadata leakage is not stated & no user study & Not stated & recovery depends on email access, fingerprint availability, and threshold shareholder cooperation, with no stated post-recovery revocation or compromise handling & hybrid wallet recovery combining social helpers, biometrics, and smart-contract coordination for self-custody key reconstruction \\
  \cite{74} & yes; public identifiers, wallet linkage, and login metadata may leak despite early privacy claims & no user study & Not stated & security and privacy depend heavily on underlying factors and public-metadata handling, with no full usability or formal security evaluation & direct non-custodial wallet recovery by re-deriving the same private key from threshold multi-factor inputs \\
  \cite{75} & encrypted transaction storage reduces account exposure, but on-chain activation can deanonymize owner & no user study; only prototype timing and cost measurements & Not stated & activation authentication is weak and recovery depends on external services plus user retention of backup key and emergency contract address & direct example of self-custody recovery by asset migration rather than key reconstruction \\
  \cite{76} & endpoint artifacts can reveal seed phrases, passwords, addresses, backups, and user activity & no evaluation & Not stated & relies on endpoint leakage and limited wallet classes; no formal security/privacy model & forensic recovery of wallet artifacts enabling investigator access or seizure, not user-facing recovery \\
  \cite{79} & protects partial data from plaintext disclosure during validation and rebuilding; metadata privacy Not stated & no end-user study; user must choose thresholds t and u & Not stated & user recovery authorization and collusion threat model are underspecified in the supplied evidence & direct wallet private-key recovery with proactive share repair and reliability analysis \\
  \cite{80} & Not stated; no explicit privacy analysis of server metadata, recovery events, or wallet linkage & no evaluation & Not stated & strong trust/availability assumptions and lack of formal or empirical validation & mobile wallet recovery architecture that reconstructs lost mnemonic shares using Shamir sharing, SIM-based 2FA, and server-assisted backup \\
  \cite{81} & helper data should not reveal original biometric vector; helper-data-only code generation probability reported below 2\textasciicircum{}-300 and 2\textasciicircum{}-246 on tested datasets & smartphone authentication around 400 ms; biometrics positioned as easier than PIN/token methods; no user study & Not stated & narrow threat model and vague financial deployment assumptions, especially around helper-data integrity, spoofing/coercion, and compromise recovery & direct example of biometric key regeneration for blockchain key-loss recovery without third-party custody \\
  \cite{82} & claims encrypted communication and unlinkability between user and guardians, but no formal privacy analysis stated & no user study & biometric encryption is described as post-quantum in passing, but no broader post-quantum recovery analysis is stated & no formal threat model or proof, and guardian identity-proof/biometric security details are underspecified & clear example of smart-contract-assisted social recovery that restores asset control by reconstructing a separate recovery key rather than the original wallet key \\
  \cite{83} & commitment hiding claimed; broader metadata privacy not stated & no user study & Not stated & not a direct end-user financial recovery workflow and relies on strong setup and coordination assumptions & transferable dynamic committee and proactive refresh pattern for private-key recoverability \\
  \cite{84} & limited; NIZK used but no explicit privacy formalization & one hardware-token operation claimed; benchmark latency reported; no user study & Not stated & no explicit treatment of hardware-token loss, client compromise, or post-recovery revocation/rotation & direct wallet private-key recovery via threshold server-held shares after MFA \\
  \cite{85} & purpose-specific and pseudonymous keys reduce linkability, but recovery-process privacy is not formalized & no user study; usability is asserted via single master-seed backup & future work mentions exploring post-quantum cryptographic algorithms & recovery authorization and root-of-trust disaster recovery are underspecified, and there is no direct validation & transferable HSM-assisted deterministic root-secret recovery pattern for managed identity and financial key infrastructures \\
  \cite{86} & claims no secret information revealed to contract, but no formal privacy model & no user study; usability argued from passkey familiarity and two-step recovery flow & Not stated & security depends on ARC storage and passkey/TK security; if ARC and TK are both compromised after registration, attacker can transfer ownership & clear example of account-control recovery for hardware-wallet users without seed reconstruction, using passkey authorization and smart-contract owner transfer \\
  \cite{87} & heuristic only; hash obfuscation and address rotation aim to reduce linkage, but no formal privacy analysis stated & no user study; usability claims only & Not stated & recovers account ownership rather than the original seed/private key, and relies on strong user artifact-storage and timing-management assumptions without formal security or usability validation & clear example of non-backup hardware-wallet recovery via on-chain ownership transfer rather than secret reconstruction \\
  \cite{89} & Not stated; no formal privacy analysis of participant or recovery metadata leakage & no evaluation & Not stated & mixed and partly vague trust assumptions plus no explicit privacy or compromise model & direct wallet private-key escrow and recovery mechanism for public blockchains with threshold reconstruction and incentive design \\
  \cite{90} & Not stated; no explicit privacy model for share metadata or recovery visibility & no evaluation & Not stated & no explicit privacy or compromise model and no post-recovery key rotation or re-sharing procedure stated & direct example of consortium-chain wallet original private key recovery via threshold secret reconstruction with incentive-based liveness support \\
  \cite{93} & formal unlinkability of session public keys; randomized basis hides information about original secret basis & no user study & yes; lattice-based construction with QROM proofs targeting post-quantum wallet security & larger signature and transaction size due to trapdoor quality degradation from lattice basis delegation & seed-based wallet key restoration in a post-quantum hot/cold self-custody design, but not a full compromise-remediation or social recovery system \\
  \cite{94} & limited for recovery workflow; broader SSI privacy claims plus masking to hide helper shares from user & no user study & not post-quantum secure; authors note future lattice/quantum direction & strong helper trust and availability assumptions with no explicit collusion or unauthorized-recovery analysis & concrete DID wallet share-recovery protocol that restores threshold signing participation without changing public key \\
  \cite{95} & helper data claimed public under fuzzy extractor assumptions; biometric information is the main secret; formal privacy analysis not stated & no user study; prototype reports about 200 ms encryption/decryption and successful recovery after 6 months & Not stated & informal security analysis with no stated revocation or compromise-recovery path for biometric leakage & direct wallet mnemonic backup and recovery mechanism using biometric-derived decryption of externally stored backup artifacts \\
  \cite{96} & zero-knowledge for backup key, but no formal privacy treatment of published x,y or replay/linkability & simplified single backup key for multiple wallets is argued, but no user study & explicit; proof and fallback-signing direction target post-quantum ownership continuity, but fallback QROM proof is missing & not true key recovery; unresolved clone attack; incomplete practical revocation and fallback-signature proof & recovery-adjacent mechanism for restoring wallet ownership and migration after key compromise \\
  \cite{97} & limited informal discussion only; no formal privacy analysis of ledger-stored seed phrases & benchmark only; no user study & Not stated & vague authorization and incomplete threat/privacy analysis for storing seed phrases in a Recovery Ledger & concrete wallet recovery design restoring seed-based wallet access, but with limited validation and underspecified trust assumptions \\
  \cite{98} & identity IDs and blockchain-published verification may expose participant and recovery metadata; privacy not formally analyzed & no evaluation & Not stated & no formal security proof and no clear recovery authorization or compromise-handling model & direct private-key recovery mechanism transferable to blockchain wallet recovery, but not tied to a concrete financial deployment \\
  \cite{99} & prevents password-equality testing across honest users; no full privacy formalization stated in evidence & no user study & efficient quantum-safe OPRF remains open & strongest security depends on advanced OPRF-based design and strong HSM/attestation assumptions; simpler variants fail under full HSM compromise & transferable password-based key recovery pattern for hardware-backed wallet or custody recovery architectures \\
  \cite{101} & qualitative only; CCTV, memo/tag tracking, and physical observation can leak sensitive information & prototype timing only; KeyGen time-consuming and Sign initially over 1 hour then about 30 minutes & Not stated & strong trusted-terminal assumption and mostly qualitative security validation & institutional custody recovery case study showing backup, threshold reconstruction, disaster recovery, and rekeying tradeoffs \\
  \cite{102} & yes, central focus on unlinkability and server-side account-linkage risks & no user study & yes, proposes pqCBR using Kyber-based components and post-quantum key-blinding signatures & no full implementation or user evaluation, single-backup-token model, and no explicit post-recovery revocation/audit workflow in the evidence & clear example of recovering account access by backup-authenticator-based credential rotation rather than reconstructing the original lost secret \\
  \cite{103} & sensitive key material handling acknowledged, but no formal privacy model for cloud processing or seized data exposure & prototype UI and workflow plus time-saving case studies; no user study & Not stated & no formal threat/privacy model and limited validation beyond narrow case studies; cloud deployment may be unacceptable for some agencies & forensic recovery of wallet artifacts and selective asset recovery from seized evidence, not end-user wallet recovery \\
  \cite{106} & encrypted backup and Solid access control are discussed, but privacy is not formally defined beyond secrecy/access restriction & no user study & Not stated & single point of failure at the Pod server and explicit failure to provide restore during DoS & backup-based restoration of SSI wallet contents and related keys after device loss; useful transferable pattern but not direct cryptoasset custody recovery \\
  \cite{107} & limited; no formal analysis of storage metadata, locations, or ordering leakage & no user study; only operational burden discussed and runtime benchmarks reported & Not stated & user must store decryption keys for all parts; initial seed generation/download remains vulnerable & concrete self-custody seed backup and reconstruction mechanism, but with underspecified authorization and limited security/usability evaluation \\
  \cite{108} & limited; no formal privacy analysis stated & user study & Not stated & strong usability evidence but incomplete formal security, privacy, and post-recovery lifecycle specification & direct wallet recovery mechanism showing factor-based restoration of wallet access without custodial key storage \\
  \cite{109} & privacy largely out of scope; no privacy formalization evidenced & no user study; usability discussed analytically through complexity, parameter choice, and response burden & Not stated & rational-adversary and static-parameter model with no real-world attack-data or deployment validation & strong for analyzing composition of wallet recovery paths and distinguishing secret reconstruction from account-control recovery \\
  \cite{111} & helper key should not reveal signing key or biometric information, but broader metadata leakage is not stated & no user study & Not stated & security theorem is limited to single-key registration, while backup/inheritance create multiple related helper keys & wallet recovery restores signing authority via helper-key regeneration rather than original-key reconstruction, with important provider-trust and biometric tradeoffs \\
  \cite{112} & template-less biometric claims, but recovery-metadata and helper-exposure privacy are not fully stated & benchmark timings and FRR/FAR experiments; no user study & uses CRYSTALS Dilithium / ML-DSA44 in implementation context & recovery depends on keystore release of factor material and uses an acknowledged weak XOR composition with no formal proof in supplied evidence & concrete same-key wallet recovery mechanism after factor loss, with keystore-assisted MFA reconstruction rather than provider custody or account transfer \\
  \cite{113} & conditional anonymity with TA traceability; committee does not learn real identity at join time; flow-analysis and network-layer privacy are out of scope & no user study; only design claims about identity-based addressing and re-registration recovery & Not stated & strong trust in TA and incomplete treatment of real-world privacy, compromise abuse, and deployment & regulated payment-system private-key recovery via identity-based reissuance, combining recovery with anonymity and traceability \\
  \cite{114} & identity-linked registration required; no formal privacy analysis & benchmark only; no user study & Not stated & reliance on trusted third-party RSP and incomplete handling of stale UTXOs/fee drift & core example of asset recovery by pre-signed transfer rather than key reconstruction \\
  \cite{117} & metadata-only audit claimed, but TTP still learns hashes, timestamps, RP identifiers, and device cluster IDs & no user study; only automation claims and latency benchmarks & Not stated & recovery semantics and collusion assumptions are unclear; evidence shows threshold signing continuity more clearly than full secret restoration & transferable threshold recovery and audit pattern for DID-linked financial access, but not an explicit asset-custody paper \\
  \cite{118} & yes; focuses on unlinkability via fresh hardened per-dApp keys, but no formal privacy proof and limited metadata-leakage discussion & prototype UI and runtime benchmarks only; no user study & Not stated & recovery still depends on retained mnemonic plus password and strong endpoint-security assumptions; no clear compromise-recovery or post-recovery revocation model & concrete non-custodial wallet recovery design combining deterministic root regeneration, encrypted decentralized backup, and smart-contract-mediated device management \\
\end{longtable}
\endgroup

\end{landscape}

\section{Seven-Stage Recovery Walkthroughs}
\label{app:seven-stage-walkthroughs}

Table~\ref{tab:seven-stage-walkthroughs} applies the construction model from Section~\ref{sec:construction-model} to four representative retained systems. The examples were selected to cover the main recovery promises requested by the model: threshold secret reconstruction, smart-account or access-structure control transfer, biometric/seed restoration, and institutional custody recovery. The comparison shows that the same seven stages produce different post-recovery obligations depending on whether the design restores the old secret, migrates control, regenerates seed material, or restores operational custody.

\begin{landscape}
\begingroup
\scriptsize
\setlength{\tabcolsep}{2pt}
\renewcommand{\arraystretch}{1.08}

\begin{longtable}{@{}L{0.11\linewidth}L{0.10\linewidth}L{0.11\linewidth}L{0.12\linewidth}L{0.11\linewidth}L{0.13\linewidth}L{0.10\linewidth}L{0.14\linewidth}@{}}
  \caption{Representative systems walked through the seven-stage construction model.}\label{tab:seven-stage-walkthroughs}\\
  \hline
  System and axis & 1. Control material & 2. Enrollment & 3. Storage and governance & 4. Failure trigger & 5. Authorization and checks & 6. Recovery action & 7. Post-recovery hygiene gap \\
  \hline
  \endfirsthead
  \caption[]{Representative systems walked through the seven-stage construction model. (continued)}\\
  \hline
  System and axis & 1. Control material & 2. Enrollment & 3. Storage and governance & 4. Failure trigger & 5. Authorization and checks & 6. Recovery action & 7. Post-recovery hygiene gap \\
  \hline
  \endhead
  \hline
  \endfoot
  PPSS wallet backup \cite{2}; secret-restoring threshold recovery & Original wallet private key or wallet-control secret & The user protects the secret with password-protected secret sharing across recovery servers & Server-held protected material plus user password; security assumes bounded server compromise and authenticated setup & Local secret availability is lost while the user still remembers the password & The claimant proves password knowledge to enough servers through the recovery protocol & The same wallet secret is reconstructed for continued use & If loss might also imply compromise, the paper leaves rotation, fund migration, stale server records, and recovery-attempt metadata handling mostly outside the core protocol \\
  Paralysis proofs \cite{10}; control-restoring access-structure transfer & Signing authority, account control, or asset-control relation under a dynamic access structure & Parties define recovery or migration rules before paralysis, including challenge-response, failed-test, or TEE-assisted paths & Governance is encoded in contracts, proofs, participants, and sometimes trusted execution; public state can expose recovery activity & A signer, custodian, or access-structure participant becomes unavailable or refuses service & A migration rule is satisfied by timeout, failed-test evidence, Merkle proof, on-chain challenge, or TEE-signed transition & Control moves to a new access structure rather than reconstructing the failed party's key & The new signer set, dispute window, failed participant status, audit trail, and old recovery path need cleanup; the old secret may remain lost or compromised \\
  Leakable mnemonic phrase \cite{95}; biometric/seed secret restoration & Wallet mnemonic or seed-equivalent backup artifact & The user enrolls biometric-derived helper data and stores encrypted mnemonic recovery material & Recovery depends on biometric recapture, helper data, external backup availability, and any optional second-factor gate & The mnemonic phrase is lost & A fresh biometric sample reproduces the decryption material through a fuzzy-extractor-style process; optional 2FA may gate helper-data download & The mnemonic or seed-equivalent material becomes available again & Restoring the same seed leaves open whether to rotate funds, replace helper data, revoke leaked biometric-derived material, or recover after biometric compromise \\
  Custodian key management \cite{101}; institutional hybrid recovery & Private keys, key shares, signing authority, and asset-transfer capability in a custody operation & Custodians create threshold shares, backup media, procedures, treasurer roles, and disaster-recovery processes & Governance sits in institutional policy, trusted terminals, M-of-N approval, physical controls, backup media, logs, and operational separation of duties & Media failure, key-management-facility disaster, personnel change, threshold change, media loss or theft, or suspected compromise & M-of-N treasurer approval and operational procedures authorize reconstruction, rekeying, or signer restoration & Operational access is restored, shares are reconstructed or refreshed, and custody control may be rekeyed & The hygiene burden is operational: rotate keys or shares after suspected compromise, update personnel and thresholds, preserve audit evidence, and review trusted-terminal and insider assumptions \\
\end{longtable}
\endgroup
\end{landscape}

\end{document}